\documentclass{article}

\usepackage{arxiv}

\usepackage[utf8]{inputenc} 
\usepackage[T1]{fontenc}    
\usepackage{hyperref}       
\usepackage{url}            
\usepackage{booktabs}       
\usepackage{amsfonts}       
\usepackage{nicefrac}       
\usepackage{microtype}      
\usepackage{lipsum}		
\usepackage{graphicx}
\usepackage{natbib}
\usepackage{doi}
\usepackage{placeins}

\usepackage{amsmath,amsfonts,amssymb,mathtools,amsthm}
\usepackage{blkarray,booktabs,multirow}
\usepackage{hyperref,cleveref}
\usepackage{graphicx,xcolor}
\usepackage[margin=0.75in,labelfont=bf,labelsep=period]{caption}
\usepackage{subfig}
\usepackage{braket}
\usepackage{arydshln}
\usepackage{mymath}
\usepackage{myqcircuit}
\usepackage{comment}
\usepackage{thmtools}
\usetikzlibrary{calc}
\usetikzlibrary{positioning}
\usetikzlibrary{automata,arrows,matrix,quotes}

\usepackage{adjustbox}
\usepackage[edges]{forest}
\usepackage{listings}

\numberwithin{equation}{section}
\newtheorem{theorem}{Theorem}[section]
\newtheorem{definition}[theorem]{Definition}

\newcommand{\rev}[1]{{#1}}
\newcommand{\revii}[1]{{#1}}

\newcommand{\opr}[1]{\ensuremath{\operatorname{#1}}}

\newcommand{\mc}[1]{\mathcal{#1}}
\newcommand{\Or}{\mathcal{O}}
\newcommand{\wt}[1]{\widetilde{#1}}
\usepackage{braket}
\usepackage{amsthm}

  \usepackage{tikz}
  \usepackage{pgfplots}
  \usetikzlibrary{calc}
  \usetikzlibrary{positioning}
  \usetikzlibrary{plotmarks}
  \usetikzlibrary{automata,arrows,matrix,quotes}
  \pgfplotsset{
    compat=newest,
    table/header=false,
    tick label style={font=\scriptsize},
    label style={font=\scriptsize},
    legend style={font=\scriptsize},
    legend cell align=left,
    colormap={parula}{
      rgb255=(53,42,135)
      rgb255=(15,92,221)
      rgb255=(18,125,216)
      rgb255=(7,156,207)
      rgb255=(21,177,180)
      rgb255=(89,189,140)
      rgb255=(165,190,107)
      rgb255=(225,185,82)
      rgb255=(252,206,46)
      rgb255=(249,251,14)
    }
  }
  \pgfplotsset{
    myColOne/.style={myblue},
    myColTwo/.style={myred},
    myColThr/.style={myorange},
    myColFou/.style={mypurple},
    myColFiv/.style={mygreen},
    myColSix/.style={mylightblue},
    myColSev/.style={mydarkred}
  }

\title{Explicit Quantum Circuits for Block Encodings of Certain Sparse Matrices}

\author{{\hspace{1mm} Daan Camps} \\
        Applied Mathematics and Computational Research Division\\ 
        Lawrence Berkeley National Laboratory \\ 
        Berkeley, CA 94720 \\
        \texttt{dcamps@lbl.gov} \\
	\And
	{\hspace{1mm}Lin Lin} \\
	Department of Mathematics and Challenge Institute of Quantum Computation\\
	University of California at Berkeley\\
        Applied Mathematics and Computational Research Division\\
        Lawrence Berkeley National Laboratory \\
	Berkeley, CA 94720 \\
	\texttt{linlin@math.berkeley.edu} \\
	\AND
        {\hspace{1mm} Roel Van Beeumen} \\
        Applied Mathematics and Computational Research Division\\ 
        Lawrence Berkeley National Laboratory \\ 
        Berkeley, CA 94720 \\
        \texttt{rvanbeeumen@lbl.gov} \\
        \AND
        {\hspace{1mm} Chao Yang} \\
        Applied Mathematics and Computational Research Division\\ 
        Lawrence Berkeley National Laboratory \\ 
        Berkeley, CA 94720 \\
        \texttt{CYang@lbl.gov} \\
}

\renewcommand{\shorttitle}{Quantum Circuits for Block Encodings of Sparse Matrices}

\hypersetup{
pdftitle={A template for the arxiv style},
pdfsubject={q-bio.NC, q-bio.QM},
pdfauthor={David S.~Hippocampus, Elias D.~Striatum},
pdfkeywords={First keyword, Second keyword, More},
}

\begin{document}
\maketitle

\begin{abstract}
Many standard linear algebra problems can be solved on a quantum computer by using recently developed quantum linear algebra algorithms that make use of block encoding\rev{s} and quantum eigenvalue / singular value transformations.  \rev{A b}lock encoding embeds a properly scaled matrix of interest $A$ in a larger unitary transformation $U$ that can be decomposed into a product of simpler unitaries and implemented efficiently on a quantum computer.  Although quantum algorithms can potentially achieve exponential speedup in solving linear algebra problems compared to the best classical algorithm, such gain in efficiency ultimately hinges on our ability to construct an efficient quantum circuit for the block encoding of $A$, which is difficult in general, and not trivial even for well structured sparse matrices. In this paper, we give a few examples on how efficient quantum circuits can be explicitly constructed  for some well structured sparse matrices, and discuss a few strategies used in these constructions.
\rev{We also provide implementations of these quantum circuits in MATLAB.}

\end{abstract}


\section{Introduction}
In recent years, a new class of quantum algorithms have been developed to solve standard linear algebra problems on quantum computers. These algorithms use 
the technique of block encoding~\cite{LowChuang2017,LowChuang2019,GilyenSuLowEtAl2019} to embed a properly scaled matrix $A$ of interest in a larger unitary matrix $U_A$ so that the multiplication of $A$ with a vector $x$ can be implemented by applying the unitary matrix $U_A$ to a carefully prepared initial state and performing measurements on a subset of qubits.
Furthermore, if $A$ is Hermitian, by using the technique of the quantum eigenvalue transformation, one can block encode a certain matrix polynomial $p(A)$ by another unitary $U_{p(A)}$ efficiently using $U_A$ as the building block~\cite{LowChuang2017}.
This procedure can be generalized to non-Hermitian matrices $A$ using a technique called the quantum singular value transformation~\cite{GilyenSuLowEtAl2019}.
The larger unitary $U_{p(A)}$ can be decomposed into a product of simpler unitaries consisting of $2\times 2$ single qubit unitaries, multi-qubit controlled-NOTs, $U_A$ and its Hermitian conjugate.  Because approximate solutions to many large-scale linear algebra problems such as linear systems of equations, least squares problems and eigenvalue problems (produced by iterative methods) can often be expressed as $p(A)v_0$ for some initial vector $v_0$, the possibility to block encode $p(A)$ will enable us to solve these problems on a quantum computer that performs unitary transformations only~\cite{GilyenSuLowEtAl2019,LinTong2020,LinTong2020a,MartynRossiTanEtAl2021}. In this paper, we use the term ``quantum linear algebra algorithms'' to refer to such block encoding based algorithms for solving linear algebra problems on a quantum computer.

In order to implement these quantum linear algebra algorithms, we need to further express the block encoding matrix $U_A$ as a product of simpler 
unitaries, i.e., we need to express $U_A$ as an efficient quantum circuit.
Although it is suggested in~\cite{Berry2015,Childs2017,LowChuang2017,GilyenSuLowEtAl2019} that such a quantum circuit can be constructed in theory for certain sparse matrices, the proposed 
construction relies on the availability of ``oracles" that can efficiently 
encode both the nonzero structure of $A$ and numerical values of the nonzero 
matrix elements.  However, for a general sparse matrix $A$, 
finding these ``oracles" is entirely non-trivial. Even for sparse matrices 
that have a well defined non-zero structure and a small set of nonzero matrix 
elements, encoding the structure and matrix elements by \rev{an} efficient quantum 
circuit is not an easy task.

In this paper, we provide a few \rev{accessible} examples on how to explicitly construct efficient quantum circuits for some well structured sparse matrices. In particular, we
will present some general strategies for writing down the oracles for 
the non-zero structure and non-zero matrix elements of these matrices. \rev{In addition to explaining the general strategies and providing circuit diagrams, we also show how these circuits can be easily constructed using a MATLAB Toolbox called \texttt{QCLAB}~\cite{qclab}. To the best of our knowledge, our work is the first to demonstrate how explicit quantum circuits can be constructed in practice to block encode structured sparse matrices.
All \texttt{QCLAB} circuit implementation can be downloaded from \url{https://github.com/QuantumComputingLab/explicit-block-encodings}.}
The block encoding of $A$ is not unique. We will show a few different 
encoding schemes and the corresponding quantum circuits. 
\rev{For a sparse matrix $A$ that has at most $s$ nonzero elements per column, which we
refer to as an $s$-sparse matrix, the general strategies we use to construct
a quantum circuit actually block encodes $A/s$. For many applications, the extra scaling factor $1/s$ does not introduce significant issues. However, for some applications such as a
quantum walk on a graph~\cite{Szegedy2004b,Childs2010}, the presence of the $1/s$ factor makes the quantum algorithm less efficient, as we will show later in this paper.
For these applications, we need to use a different strategy to block encode
$A$ directly.  We will show how this can be done for a quantum walk in which
the matrix $A$ is a symmetric stochastic matrix or a discriminant matrix, and
the corresponding quantum walk can be viewed as a block encoding of a Chebyshev 
polynomial of the discriminant matrix associated with the stochastic matrix $A$ 
used to describe the corresponding random walk. 
} 

This paper is organized as follows. In \cref{sec:blkencode}, we give a formal definition of \rev{a} block encoding of a matrix $A$ and the possibility of using the quantum eigenvalue transformation to block encode $p(A)$ for some polynomial $p(t)$.  In \cref{sec:circuit}, we discuss how to construct a block encoding circuit for a scaled $s$-sparse matrix $A/s$, and give two specific examples. 
We discuss the possibility to use the techniques presented in \cref{sec:circuit} to construct a circuit for the block encoding of $P/s$ in \cref{sec:qw}, where $P$ is a symmetric stochastic a matrix associated with a random walk. We show this circuit can be used as a building block to construct a circuit for performing a quantum walk associated with such a random walk. \rev{We explain why} such a construction leads to a loss of quantum efficiency due to the $1/s$ scaling factor produced in the block encoding schemes used in \cref{sec:circuit}. An alternative approach that takes advantage of the stochastic property of a random walk and block encodes $P$ directly instead of $P/s$ is presented. As a result, we can construct an  efficient block encoding for a Chebyshev polynomial of $P$. \rev{The block encoding view of a quantum walk introduced in this section differs from how quantum walks (in particular the \rev{Szegedy quantum walks~\cite{Szegedy2004b}}) are traditionally introduced. We make a clear connection between these two views in \cref{sec:qw} and show in the supplementary materials how the block encoding view can be used to explain the efficiency of a quantum walk compared \revii{to} a classical random walk.}

%
%
%

\section{Notations and Conventions}
Following standard conventions used in the quantum computing literature, we use the Dirac $\bra{\cdot}$ and $\ket{\cdot}$ notation to denote respectively row and column vectors.  In particular
$\ket{0}$ and $\ket{1}$ are used to represent \rev{the unit} vectors $e_{\rev{0}} = [1 \ 0]^T$ and $e_{\rev{1}} = [0 \ 1]^T$, respectively.
The tensor product of $m$ $\ket{0}$'s is denoted by $\ket{0^m}$. We use $\ket{x,y}$ to represent the Kronecker product of $\ket{x}$ and $\ket{y}$, which is also sometimes written as $\ket{x}\ket{y}$ or $\ket{xy}$.
The $N\times N$ identity matrix is denoted by $I_N$ and we sometimes drop the subscript $N$
when the dimension is clear from the context.
The $j$th column of $I_N$ is denoted by $\ket{j}$ for $j=0,1,2...,N-1$.
\rev{For the binary representation of $j \inN: 0 \leq j \leq 2^n-1$, we use the \emph{little-endian} convention as given by
\[
j = [\ttj_{n-1} \cdots \ttj_1 \ttj_0]
  = \ttj_{n-1} \cdot 2^{n-1} + \cdots + \ttj_1 \cdot 2^1 + \ttj_0 \cdot 2^0,
\]
where $\ttj_i \in \lbrace 0,1 \rbrace$ for $i = 0,\ldots,n-1$.}

We use the letters $H$, $X$\rev{, $Y$,} and $Z$ to represent the Hadamard, Pauli-$X$\rev{, Pauli-$Y$,} and Pauli-$Z$ matrices, respectively, defined below
\begin{align}
H &= \frac{1}{\sqrt{2}} \begin{bmatrix}
1 & \phantom{-}1 \\
1 & -1
\end{bmatrix}, &
X &= \begin{bmatrix}
0 & 1 \\
1 & 0
\end{bmatrix}, &
Y &= \begin{bmatrix}
0 & -i \\
i &  \phantom{-}0     
\end{bmatrix}, &
Z &= \begin{bmatrix}
1 &  \phantom{-}0 \\
0 & -1 
\end{bmatrix}.
\label{eq:basicgates}
\end{align}
\rev{Further, rotation matrices along the Pauli-$Y$ axis will play a key role in the construction of block encoding circuits and are defined as follows}
\begin{equation}
\rev{R_y(\theta) := \begin{bmatrix}
\cos\left(\frac\theta2\right) & -\sin\left(\frac\theta2\right) \\[3pt]
\sin\left(\frac\theta2\right) & \phantom{-}\cos\left(\frac\theta2\right)
\end{bmatrix}
= e^{-i\theta Y/2},}
\label{eq:def-Ry}
\end{equation}
\rev{where $\theta$ is the rotation angle and the unitary $Y$ defined in \eqref{eq:basicgates}.}
The matrices in \cref{eq:basicgates,eq:def-Ry} are $2\times 2$ unitaries and serve as basic single qubit quantum gates.

\begin{figure}[hbtp]
\centering
\subfloat[\label{fig:circuit1}]{%
\begin{small}
\begin{tikzpicture}[on grid]
\pgfkeys{/myqcircuit, layer width=7.5mm, row sep=5mm, source node=qwsource}
\newcommand{\qwstart}{1}
\newcommand{\qwend}{3}
\qwire[index=1, start layer=\qwstart, end layer=\qwend, label=$q_0$]
\qwire[index=2, start layer=\qwstart, end layer=\qwend, label=$q_1$]
\qwire[index=3, start layer=\qwstart, end layer=\qwend, label=$q_2$]
\multiqubit[layer=1, start index=1, stop index=3, label=$U$]
\end{tikzpicture}%
\end{small}%
}
%
\quad \quad
\subfloat[\label{fig:circuit2}]{%
\begin{small}
\begin{tikzpicture}[on grid]
\pgfkeys{/myqcircuit, layer width=7.5mm, row sep=5mm, source node=qwsource}
\newcommand{\qwstart}{1}
\newcommand{\qwend}{3}
\qwire[index=1, start layer=\qwstart, end layer=\qwend, label=$\ket{j_2}$]
\qwire[index=2, start layer=\qwstart, end layer=\qwend, label=$\ket{j_1}$, start node=qws]
\qwire[index=3, start layer=\qwstart, end layer=\qwend, label=$\ket{j_0}$]
\multiqubit[layer=1, start index=1, stop index=3, label=$U$]
\node[left] at ($(qws)-(3ex,0ex)$) {$\ket{j}\left\{\phantom{\begin{matrix}~\\[3pt]~\\[3pt]~\end{matrix}}\right.$};
\end{tikzpicture}%
\end{small}%
}
\caption{(a) Qubits of a quantum circuit $U$ are numbered increasingly from the top to the bottom. (b) How the binary representation of an integer ($j$) input is mapped to the qubits of the quantum circuit $U$.  }
\label{fig:circuit}
\end{figure}
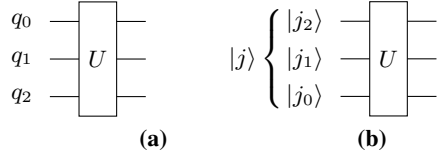

We follow the standard convention for drawing quantum circuits with multiple parallel lines
covered by several layers of rectangular boxes. Each line corresponds to either a single qubit or
multiple qubits depending on how it is labelled and
each box corresponds to a single qubit or multi-qubit gate \rev{depending on the number of qubit lines passing through it.
We use the convention that
the qubits in a circuit diagram  are numbered increasingly from the top to the bottom as illustrated by the 3 qubit circuit $U$ in \cref{fig:circuit1}. An integer $j = [\ttj_{n-1} \cdots \ttj_1 \ttj_0]$ input to a circuit is prepared as a set of quantum states $\ket{\ttj_{n-1}}$,\ldots,$\ket{\ttj_0}$ with $\ket{\ttj_0}$ mapped to the highest numbered qubit $q_{n-1}$ and $\ket{\ttj_{n-1}}$ mapped to the lowest numbered qubit $q_0$, see e.g., \cref{fig:circuit2}.}

\begin{figure}[hbtp]
\centering
\begingroup
\captionsetup[subfigure]{width=0.5\linewidth}
\subfloat[$\ket0$ controlled\label{fig:cnot0}]{%
\begin{small}
\begin{tikzpicture}[on grid]
\pgfkeys{/myqcircuit, layer width=7.5mm, row sep=5mm, source node=qwsource}
\newcommand{\qwstart}{1}
\newcommand{\qwend}{3}
\qwire[start node=qw1s, end node=qw1e, style=thin, index=1, start layer=\qwstart, end layer=\qwend, label=$q_0$]
\qwire[start node=qw2s, end node=qw2e, style=thin, index=2, start layer=\qwstart, end layer=\qwend, label=$q_1$]
\cnot[layer=1, control index=1, target index=2, style=controloff]
\node[right] at ($0.5*(qw1e)+0.5*(qw2e)$) {$=\ \begin{bmatrix}
0 & 1 & 0 & 0 \\
1 & 0 & 0 & 0 \\
0 & 0 & 1 & 0 \\
0 & 0 & 0 & 1
\end{bmatrix}$};
\end{tikzpicture}%
\end{small}%
}
\endgroup
\quad \quad
%
\begingroup
\captionsetup[subfigure]{width=0.5\linewidth}
\subfloat[$\ket1$ controlled\label{fig:cnot1}]{%
\begin{small}
\begin{tikzpicture}[on grid]
\pgfkeys{/myqcircuit, layer width=7.5mm, row sep=5mm, source node=qwsource}
\newcommand{\qwstart}{1}
\newcommand{\qwend}{3}
\qwire[start node=qw1s, end node=qw1e, index=1, start layer=\qwstart, end layer=\qwend, label=$q_0$]
\qwire[start node=qw2s, end node=qw2e, index=2, start layer=\qwstart, end layer=\qwend, label=$q_1$]
\cnot[layer=1, control index=1, target index=2]
\node[right] at ($0.5*(qw1e)+0.5*(qw2e)$) {$=\ \begin{bmatrix}
1 & 0 & 0 & 0 \\
0 & 1 & 0 & 0 \\
0 & 0 & 0 & 1 \\
0 & 0 & 1 & 0
\end{bmatrix}$};
\end{tikzpicture}%
\end{small}%
}
\endgroup
\caption{Controlled-NOT gates.}
\label{fig:cnotgate}
\end{figure}
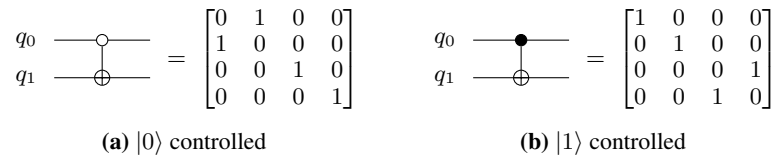

\rev{An important class of quantum gates are the controlled gates, i.e., one or more qubits act as a control for some operation.
Graphically, the control operation is represented by a vertical line connecting the control qubit(s), marked by either a solid or open circle, to the so called target gate, see e.g., the controlled NOT (CNOT) in \cref{fig:cnotgate}.  The target can also be multiple gates grouped in a subcircuit block used to perform certain operations.
A solid circle indicates that the controlled operation is performed on the connected qubit, when the input to the controlling qubit is a $\ket{1}$ state.
Similarly, an open circle indicates that the controlled operation is performed when the input to the controlling qubit is a $\ket{0}$ state.
}

Mathematically, the controlled gates from \cref{fig:cnotgate} translate into
\begin{equation}
\rev{E_0 \otimes X + (I - E_0) \otimes I},
\qquad \mbox{and} \qquad
\rev{E_1 \otimes X + (I - E_1) \otimes I},
\end{equation}
respectively, where \rev{the orthogonal projection operators}
\begin{align}
E_{\rev{0}} &= e_{\rev{0}} e_{\rev{0}}^T = \ket{0}\bra{0}, &
E_{\rev{1}} &= e_{\rev{1}} e_{\rev{1}}^T = \ket{1}\bra{1},
\label{eq:e1e2}
\end{align}
\rev{serve as the controls. Note that the NOT ($X$) operation is applied to the input of qubit $q_1$ in \cref{fig:cnot0} only if the input to qubit $q_0$ is $\ket{0}$. Likewise, the NOT operation is applied to the input of qubit $q_1$ in \cref{fig:cnot1} if the input to qubit $q_0$ is $\ket{1}$.} A similar expression can be
use to denote multi-qubit controlled NOT gates.

\rev{Throughout this paper, we use the \texttt{QCLAB} Toolbox\footnote{\texttt{QCLAB} Toolbox: \url{https://github.com/QuantumComputingLab/qclab}} to construct and draw quantum circuits for block encodings of matrices in \textsc{MATLAB}.
Hence, it will be convenient to verify that a presented quantum circuit performs the correct unitary transformation.
As a simple example, the following listing constructs a simple quantum circuit with $n = 2$ qubits that prepares a so-called Bell state when the input of the circuit is $\ket{00}$.  The circuit consists of a Hadamard gate on the first qubit, followed by a CNOT gate with control on the first and target on the second qubit.}

\begin{minipage}{0.675\textwidth}
\begin{lstlisting}[language=matlab,basicstyle=\ttfamily\small,keywordstyle=\color{blue},commentstyle=\color{green!70!black}]
circuit = qclab.QCircuit(2);
circuit.push_back(qclab.qgates.Hadamard(0));
circuit.push_back(qclab.qgates.CNOT(0,1));
\end{lstlisting}
\end{minipage}%
\hfill%
\begin{minipage}{0.225\textwidth}
\begin{small}
\begin{tikzpicture}[on grid]
\pgfkeys{/myqcircuit, layer width=7.5mm, row sep=5mm, source node=qwsource}
\newcommand{\qwstart}{1}
\newcommand{\qwend}{4}
\qwire[start node=qw1s, end node=qw1e, style=thin, index=1, start layer=\qwstart, end layer=\qwend, label=$q_0$]
\qwire[start node=qw2s, end node=qw2e, style=thin, index=2, start layer=\qwstart, end layer=\qwend, label=$q_1$]
\singlequbit[style=gate, layer=1, index=1, node=H1, label=$H$]
\cnot[layer=2, control index=1, target index=2]
\end{tikzpicture}%
\end{small}
\end{minipage}

\noindent
\rev{%
\texttt{QCLAB} uses MATLAB's object oriented programming capabilities to keep track of all gates used in the circuit and to which qubits these gates are applied to and in what order.  The first line of the listing defines a \texttt{QCLAB} circuit object \texttt{circuit} consisting of 2 qubits.  The circuit is constructed by using the \texttt{push\_back} function to place quantum gates (\texttt{qgates}) one layer at a time into the \texttt{circuit} object  according to the circuit diagram shown to the right of the listing.
The unitary matrix defined by the circuit is not explicitly formed unless the \texttt{circuit.matrix} function is called. One can also draw a circuit by using the \texttt{circuit.draw} function. We refer readers to the User's manual of \texttt{QCLAB}~\cite{qclab} for the syntax of its usage.}


\section{Block encoding and quantum eigenvalue transformation}
\label{sec:blkencode}
Block encoding is a technique for embedding a properly scaled nonunitary matrix $A \in \mathbb{C}^{N\times N}$ into a unitary matrix $U_A$ of the form
\begin{equation}
U_A = \begin{bmatrix}
A & \ast \\
\ast & \ast
\end{bmatrix},
\end{equation}
where $\ast$ denotes a matrix block yet to be determined.
Applying $U_A$ to a vector of the form
\begin{equation}
v = \begin{bmatrix}
x \\
0
\end{bmatrix} = \ket{0}\ket{x},
\end{equation}
\rev{where $\|x\|_2=1$}, yields
\[
w = U_A v = \begin{bmatrix}
Ax \\
*
\end{bmatrix} = \ket{0}(A\ket{x}) + \ket{1}\ket{\ast},
\]
where $\ast$ here denotes a vector that we do not care about.
If we measure the first qubit and obtain the $\ket{0}$ state, the
second qubit register then contains $A\ket{x}$. The probability of such a 
successful measurement is $\|Ax\|^2$.   

Note that by ``a properly scaled $A$", we mean that $A$ is scaled to satisfy 
$\|A\|_2 \le 1$. Without such a scaling, a block encoding of $A$ may not exist \rev{because the singular values of any submatrix blocks of a unitary matrix must be bounded by $1$.}
Furthermore, if there are some constraints on the type of unitary we can
construct, e.g., the type of quantum gates available on a quantum computer, we may not be able to find a $U_A$ that block encodes $A$ exactly.  
However, in this paper, we assume that $A$ has already been properly scaled, and
there is no constraints on the quantum gates we can use to construct
a quantum circuit representation of the unitary matrix $U_A$.

To give an example of a block encoding, let us consider a $1\times 1$ matrix 
$A = \alpha$, where $0 < \alpha < 1$. In this extremely simple case, a 
block encoding of $A$ can be constructed as 
\begin{equation}
U_{A} = \begin{bmatrix}
\alpha & \sqrt{1-\alpha^2} \\
\sqrt{1-\alpha^2} & -\alpha
\end{bmatrix},
\qquad \mbox{or} \qquad
U_{A} = \begin{bmatrix}
\alpha & -\sqrt{1-\alpha^2} \\
\sqrt{1-\alpha^2} & \alpha
\end{bmatrix}.
\label{eq:bealpha}
\end{equation}
Although this type of block encoding can be extended to a properly scaled matrix
$A$ of a larger dimension to yield
\begin{equation}
U_{A} = \begin{bmatrix}
 A & (I-A^{\dagger}A)^{1/2} \\
(I-A^{\dagger}A)^{1/2} & -A
\end{bmatrix},
\ \ \mbox{or} \ \
\begin{bmatrix}
A & -(I-A^{\dagger}A)^{1/2} \\
(I-A^{\dagger}A)^{1/2} & A
\end{bmatrix},
\label{eq:sqrtbe}
\end{equation}
this approach is not practical because it requires computing the square root of $A^{\dagger}A$, which \rev{requires computing and diagonalizing $A^{\dagger}A$. In general, there is no efficient algorithm to perform these operations on a quantum computer using $\mathcal{O}(\mathrm{poly}(n))$ quantum gates.}

A more practical scheme that does not require computing the square root of $A$ can be illustrated by the following real symmetric $2\times 2$ example. Let
\begin{equation}
A = \begin{bmatrix}
\alpha_{1}  & \alpha_{2} \\
\alpha_{2}  & \alpha_{1}
\end{bmatrix},
\label{eq:a2}
\end{equation}
where $|\alpha_1|, |\alpha_2| \leq 1$.
It can be verified that the matrix
\begin{equation}
U_{A} = \frac{1}{2} \begin{bmatrix}
U_{\alpha} & -U_{\beta} \\
U_{\beta} &  U_{\alpha}
\end{bmatrix}
\label{eq:UAB}
\end{equation}
is a block encoding of $A/2$, where
\begin{align}
U_{\alpha} &=\begin{bmatrix}
 \alpha_1 &  \alpha_2 &  \alpha_1  &  -\alpha_2 \\
 \alpha_2 &  \alpha_1 & -\alpha_2  &   \alpha_1 \\
 \alpha_1 & -\alpha_2 &  \alpha_1  &   \alpha_2 \\
-\alpha_2 &  \alpha_1 &  \alpha_2  &   \alpha_1
\end{bmatrix}, &
U_{\beta} &= \begin{bmatrix}
 \beta_1 &  \beta_2 &  \beta_1  &  -\beta_2 \\
 \beta_2 &  \beta_1 & -\beta_2  &   \beta_1 \\
 \beta_1 & -\beta_2 &  \beta_1  &   \beta_2 \\
-\beta_2 &  \beta_1 &  \beta_2  &   \beta_1
\end{bmatrix},
\label{eq:UAB2}
\end{align}
with $\beta_1 = \sqrt{1-\alpha_1^2}$ and $\beta_2 = \sqrt{1-\alpha_2^2}$.
To implement this unitary on a quantum computer, we must further decompose $U_A$ as
a product of simpler unitaries and construct a quantum circuit with a limited
number of quantum gates.  We will discuss how to construct such a quantum
circuit in the next section.
The method for constructing the block encoding shown in \eqref{eq:UAB}--\eqref{eq:UAB2} cannot be easily generalized to matrices of larger dimensions. However, for matrices with special structures, e.g., sparse 
matrices, it is possible to develop some general block encoding strategies which we will discuss in the next section.
The general definition of a block encoding is as follows.

\begin{definition}[Block encoding] Given an $n$-qubit matrix $A$ ($N=2^n$), if we can find $\alpha, \epsilon \in \mathbb{R}_+$, and an $(m+n)$-qubit unitary matrix $U_A$ so that 
\begin{equation}
\Vert A - \alpha \left(\langle 0^m | \otimes I_N\right) U_A \left( | 0^m \rangle \otimes I_N \right) \Vert_2 \leq \epsilon,
\end{equation}
then $U_A$ is called an $(\alpha, m, \epsilon)$-block-encoding of $A$.
In particular, when the block encoding is exact with $\epsilon=0$, $U_A$ is called an $(\alpha, m)$-block-encoding of $A$. 
\label{def:blockencode}
\end{definition}

Here $m$ is the number of ancilla qubits used to block encode $A$, and the expression $\left(\langle 0^m | \otimes I_N\right) U_A \left( | 0^m \rangle \otimes I_N \right)$ should be interpreted as taking the upper-left $2^n\times 2^n$ matrix block of $U_A$. For example, the block encoding in \eqref{eq:UAB}--\eqref{eq:UAB2} is a $(1,2)$-block-encoding of $A$ in \cref{eq:a2}.

When solving a large and sparse linear algebra problem on a classical computer, we often
use an iterative solver to seek an approximate solution $\hat{x}$.  For simplicity we assume $A$ is Hermitian. In many cases, such as when a Krylov subspace method is used, the approximate solution can be written as $\hat{x}=p(A)v_0$, where $p(t)$ is a polynomial that approximates a desired function and $v_0$ is an initial guess or simply a random vector.  
For example, to solve a linear system of equations $Ax=b$, we choose $p(t)$ to be a polynomial approximation to $1/t$ on the spectrum of $A$ and $v_0=b$. 

\begin{figure}[!ht]
\centering
\begin{small}
\begin{tikzpicture}[on grid]
\pgfkeys{/myqcircuit, layer width=8mm, row sep=5mm, source node=qwsource}
\newcommand{\qwstart}{1}
\newcommand{\qwend}{11}
\qwire[start node=qw1s, end node=qw1e, style=thin, index=1, start layer=\qwstart, end layer=\qwend, label=$\ket{0}$]
\qwire[start node=qw2s, end node=qw2e, style=thin, index=2, start layer=\qwstart, end layer=\qwend, label=$\ket{0^m}$]
\qwire[start node=qw3s, end node=qw3e, style=thin, index=3, start layer=\qwstart, end layer=\qwend, label=$\ket{\psi}$]

\singlequbit[style=gate, layer=1, index=1, node=H1, label=$H$]
\singlequbit[style=not, layer=2, index=1, node=not1]
\control[layer=2, index=2, target node=not1, style=controloff]
\singlequbit[style=gate, layer=3, index=1, node=R2d, label=$e^{-i\phi_{d}Z}$]
\singlequbit[style=not, layer=4, index=1, node=not2]
\control[layer=4, index=2, target node=not2, style=controloff]
\multiqubit[layer=5, start index=2, stop index=3, label=$U_A$]
\singlequbit[style=not, layer=6, index=1, node=not3]
\control[layer=6, index=2, target node=not3, style=controloff]
\singlequbit[style=gate, layer=7, index=1, node=Rd, label=$e^{-i\phi_{d-1}Z}$]
\singlequbit[style=not, layer=8, index=1, node=not4]
\control[layer=8, index=2, target node=not4, style=controloff]
\multiqubit[layer=9, start index=2, stop index=3, label=$U_A^{\dagger}$]

\pgfkeys{/myqcircuit, gate offset=1}
\renewcommand{\qwstart}{12}
\renewcommand{\qwend}{16}

\qwire[start node=qw1s, end node=qw1e, style=thin, index=1, start layer=\qwstart, end layer=\qwend, label={$\cdots$}]
\qwire[start node=qw2s, end node=qw2e, style=thin, index=2, start layer=\qwstart, end layer=\qwend, label={$\cdots$}]
\qwire[start node=qw3s, end node=qw3e, style=thin, index=3, start layer=\qwstart, end layer=\qwend, label={$\cdots$}]

\singlequbit[style=not, layer=10, index=1, node=not5]
\control[layer=10, index=2, target node=not5, style=controloff]
\singlequbit[style=gate, layer=11, index=1, node=R0, label=$e^{-i\phi_{0}Z}$]
\singlequbit[style=not, layer=12, index=1, node=not6]
\control[layer=12, index=2, target node=not6, style=controloff]
\singlequbit[style=gate, layer=13, index=1, node=H2, label=$H$]


\end{tikzpicture}%
\end{small}%
\caption{A quantum circuit for the block encoding of $p(A)$.}
\label{fig:qetcirc}
\end{figure}
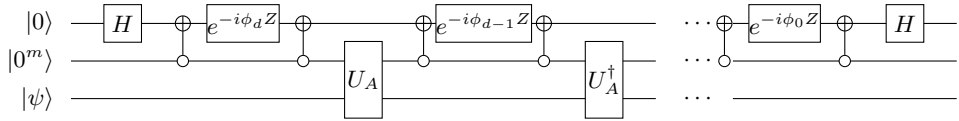

To solve this problem on a quantum computer using the technique of block encoding, we need to block encode the matrix function $p(A)$. 
If we are provided a $(1,m)$-block-encoding of the Hermitian matrix $A$ denoted by $U_A$, this task can be achieved by a quantum eigenvalue transformation \rev{(QET)} \cite{LowChuang2017,GilyenSuLowEtAl2019}.
Specifically, for any real polynomial $p(t)$ of degree $d$ satisfying (i) the parity of $p$ is $(d \bmod 2)$, and (ii) $|p(t)|\le 1, \forall t \in [-1, 1]$, there exists a set of parameters $\{\phi_i\}_{i=0}^d\in\mathbb{R}^{d+1}$ so that the quantum circuit in \cref{fig:qetcirc} provides a $(1,m+1)$-block-encoding of $p(A)$~\cite[Corollary 11]{GilyenSuLowEtAl2019} denoted by $U_{p(A)}$.  

\rev{The QET theory and the corresponding quantum circuit for block encoding
a polynomial of $A$ in terms of the block encoding of $A$ is a consequence of  quantum signal processing (QSP)~\cite{LowChuang2017,GilyenSuLowEtAl2019}, which represents a scalar polynomial of degree $d$ using a product of unitary matrices of size $2\times 2$, parameterized by $(d+1)$ real numbers called the phase factors.
We state the result of QSP below for completeness, which is a slight variation of \cite[Theorem 4]{GilyenSuLowEtAl2019}.
\begin{theorem}[Quantum signal processing] \label{thm:qsp}
Let 
\begin{equation}
 U(t) = 
 \begin{bmatrix}
 t & \sqrt{1-t^2} \\
 \sqrt{1-t^2} & -t
 \end{bmatrix}.
\label{eq:Ut}
\end{equation}
There exists a set of phase angles $\Phi_d \equiv \{\phi_0,,...,\phi_d\} \in \mathbb{R}^{d+1}$so that 
\begin{equation}
U_{\Phi_d}(t) \equiv  e^{i\phi_0 Z} \Pi_{j=1}^d \left[U(t)e^{i\phi_j Z} \right] 
 =
 \begin{bmatrix}
 p(t) & -q(t)\sqrt{1-t^2} \\
 q^{\ast}(t)\sqrt{1-t^2} & p^{\ast}(t)
 \end{bmatrix},
 \label{eq:Uphi}
\end{equation}
if and only if $p(t)$ and $q(t)$ are complex valued polynomials in $t$ and satisfy
\begin{enumerate}
\item $\mathrm{deg}(p) \leq d$, $\mathrm{deg}(q) \leq d-1$;
\item $p$ has parity $d$ mod $2$ and $q$ has parity $d-1$ mod $2$;
\item $|p(t)|^2 + (1-t^2)|q(t)|^2 = 1$, $\forall t\in [-1,1]$.
\end{enumerate}
When $d=0$, $\deg(q)\le-1$ should be interpreted as $q=0$.
\end{theorem}
}
 
\rev{
We may verify that if $\Phi_d \equiv (\phi_0,\phi_1,...,\phi_d)$ is 
chosen as
\[
\phi_j = \left\{
\begin{array}{ll}
\pi/4, & j = 0 \: \mathrm{or} \: d  \\
\pi/2, & j = 1,...,d-1,
\end{array}
\right.
\]
then $p(t)$ is the $d$th degree Chebyshev polynomial of the first kind (up to a constant global factor $i^d$). However, for a properly normalized arbitrary $d$th degree (real valued) polynomial $p(t)$, finding the phase angles in $\Phi_d \equiv (\phi_0,\phi_1,...,\phi_d)$ to make \eqref{eq:Uphi} hold is not trivial. In the past few years, there has been significant progress in developing new algorithms to improve the efficiency and robustness to find phase factors~\cite{GilyenSuLowEtAl2019,Haah2019,ChaoDingGilyenEtAl2020,DongMengWhaleyEtAl2021,Ying2022,DongLinNiEtAl2022}.}


\rev{The approach adopted in this paper is the optimization based method~\cite{DongMengWhaleyEtAl2021}, which computes the phase angles $\phi_0$, $\phi_1$, ..., $\phi_d$ by solving the nonlinear least squares problem
\[
\min_{\Phi_d} \sum_{k=1}^{\wt{d}} | \Re [e_1^T U_{\Phi_d}(t_k)e_1] - p_d(t_k) |^2,
\] 
where $t_k$ are chosen to be roots of a Chebyshev polynomial, and $\wt{d}=\lceil (d+1)/2\rceil$.
This approach is particularly useful when we are targeting a \textit{real} polynomial $p(t)$. 
With a proper choice of the initial guess, and despite the complex global optimization landscape~\cite{WangDongLin2022}, this approach can robustly find phase factors for high degree polynomials ($d\sim 10^4$).}


Once we have an efficient quantum circuit to construct the block encoding $U_A$, we can easily construct an efficient quantum circuit for
$U_{p(A)}$. For a general matrix $A$, the quantum circuit in \cref{fig:qetcirc} no longer corresponds to a QET, but a quantum singular value transformation (QSVT) of $A$ (see~\cite{GilyenSuLowEtAl2019} for more detailed discussion of QSVT).

In the next section, we will focus on techniques for constructing an efficient quantum circuit for $U_A$ associated with a well structured sparse $A$. 

\section{Efficient quantum circuits for block encodings of $s$-sparse matrices}
\label{sec:circuit}

\begin{figure}[hbtp]
\begin{minipage}{0.39\textwidth}
\begin{small}
\begin{tikzpicture}[on grid]
\pgfkeys{/myqcircuit, layer width=5mm, row sep=6mm, source node=qwsource}
\newcommand{\qwstart}{1}
\newcommand{\qwend}{9}
\qwire[start node=qw1s, end node=qw1e, style=thin, index=1, start layer=\qwstart, end layer=\qwend, label=$\ket{0}$]
\qwire[start node=qw2s, end node=qw2e, style=thin, index=2, start layer=\qwstart, end layer=\qwend, label=$\ket{0}$]
\qwire[start node=qw3s, end node=qw3e, style=thin, index=3, start layer=\qwstart, end layer=\qwend, label=$\ket{j}$]

\singlequbit[layer=1, index=2, node=H1, label=$H$]
\singlequbit[layer=2, index=1, node=R1, label=$R_1$]
\cnot[layer=3, control index=2, target index=1]
\singlequbit[layer=4, index=1, node=R2, label=$R_2$]
\cnot[layer=5, control index=2, target index=1]
\cnot[layer=5.5, control index=2, target index=3]
\singlequbit[layer=6.5, index=2, node=H2, label=$H$]
\singlequbit[style=meter, layer=7.75, index=1, node=meas]
\singlequbit[style=meter, layer=7.75, index=2, node=meas]
\end{tikzpicture}%
\end{small}
\end{minipage}\hfill%
\begin{minipage}{0.61\textwidth}
\begin{lstlisting}[language=matlab,basicstyle=\ttfamily\footnotesize]
C = qclab.QCircuit(3);
C.push_back(qclab.qgates.Hadamard(1));
C.push_back(qclab.qgates.RotationY(0,phi1));
C.push_back(qclab.qgates.CNOT(1,0));
C.push_back(qclab.qgates.RotationY(0,phi2));
C.push_back(qclab.qgates.CNOT(1,0));
C.push_back(qclab.qgates.CNOT(1,2));
C.push_back(qclab.qgates.Hadamard(1));
\end{lstlisting}
\end{minipage}
\caption{A quantum circuit for the block encoding of a $2\times 2$ symmetric matrix $A$.}
\label{fig:uhcirc1}
\end{figure}
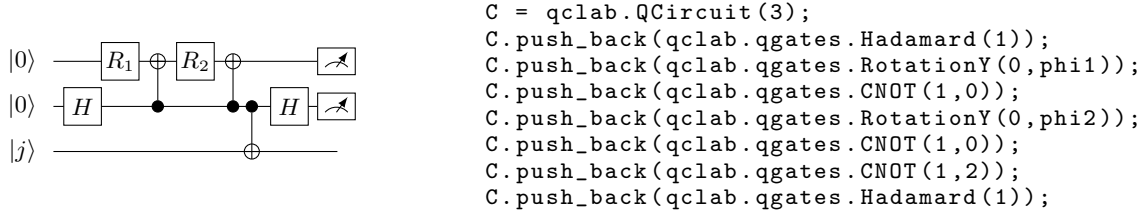

\rev{In this section, we introduce strategies to directly construct quantum circuits for block encodings of structured and sparse matrices.
Before diving into all the details, we first return to real symmetric $2 \times 2$ example of \eqref{eq:a2}.
If we define $\phi_1 = \arccos(\alpha_1) + \arccos(\alpha_2)$ and $\phi_2 = \arccos(\alpha_1) - \arccos(\alpha_2)$, then} we can verify that the block encoding \eqref{eq:UAB}--\eqref{eq:UAB2} for the $2\times 2$ matrix \eqref{eq:a2} can be factored as a product of simpler unitaries, i.e.,
\rev{
\begin{equation}
U_{A} = U_6 U_5 U_4 U_3 U_2 U_1 U_0,
\label{eq:uhfac}
\end{equation}
where
$U_0 = U_6 = I_2 \otimes H \otimes I_2$, 
$U_1 = R_1 \otimes I_2 \otimes I_2$,
$U_2 = U_4 = (I_2 \otimes E_0 + X \otimes E_1) \otimes I_2$,
$U_3 = R_2 \otimes I_2 \otimes I_2$,
and 
$U_5 = I_2 \otimes (E_0 \otimes I_2 + E_1 \otimes X)$.
}
Here, $H$, $X$ are the Hadamard and Pauli-$X$ gates defined in \eqref{eq:basicgates}, \rev{$R_1 = R_y(\phi_1)$, $R_2 = R_y(\phi_2)$ are the rotation matrices defined in \eqref{eq:def-Ry}}, and $E_0$, $E_1$ are projectors defined in \eqref{eq:e1e2}.
The quantum circuit associated with the factorization given in \eqref{eq:uhfac}
is shown in \cref{fig:uhcirc1}.

Note that in this case, the quantum circuit requires 2 ancilla qubits in addition to 
the $n = 1$ system qubit required to match the dimension of $A$, which is $N = 2^n$.
As a result, the unitary that block encodes the $2\times 2$ matrix $A$
is of dimension $2^3$, which is twice the dimension of the block
encoding given by \eqref{eq:sqrtbe}. 

\rev{Although it may not be immediately clear how \eqref{eq:uhfac} and the corresponding quantum circuit shown in \cref{fig:uhcirc1} are obtained from 
the block encoding matrix \eqref{eq:UAB}--\eqref{eq:UAB2}, it is possible, as we will show below, to develop a general scheme to construct a block encoding and the corresponding quantum circuit for well-structured matrices.  In particular, when the matrix $A$ is sparse with a structured sparsity pattern and the nonzero matrix elements can also be well characterized, an efficient block encoding circuit for $A$  can be constructed, as we will show in the next section.}

In the following, we will focus on an $s$-sparse matrix, which is defined to be an sparse matrix with at most $s$ nonzeros in each column and row. For simplicity,
let us assume $s=2^m \ll N$ for some integer $m \ll n$.  When 
$s$ is not a power of 2, we can always increase $s$ to a power of 2 by treating 
some zeros as nonzeros.
 The following theorem, which is a variant of~\cite[Lemma 48 (in the full version)]{GilyenSuLowEtAl2019}, asserts that such 
a sparse matrix can be block encoded if we can construct unitaries that can encode both
the nonzero structure and the nonzero matrix elements of $A$.

\begin{theorem} \label{th:Usparse}
Let $c(j,\ell)$ be a function that gives the \revii{row} index of the $\ell$th \revii{(among a list of $s$) non-zero matrix elements} in the $j$th column of an $s$-sparse matrix $A \in \mathbb{C}^{N\times N}$ with $N=2^n$, where $s=2^m$.  If there exists a unitary $O_c$ such that 
\begin{equation}
O_c \ket{\ell}\ket{j} = \ket{\ell}\ket{c(j,\ell)},
\label{eq:oc}
\end{equation}
and a unitary $O_A$ such that
\begin{equation}
O_A \ket{0}\ket{\ell}\ket{j} = \left(
A_{c(j,\ell),j} \ket{0} + \sqrt{1-|A_{c(j,\ell),j}|^2} \ket{1}
\right)
\ket{\ell}\ket{j},
\label{eq:oa}
\end{equation}
then 
\begin{equation}
U_A = \left(I_2 \otimes D_s\otimes I_N \right) 
\left(I_2 \otimes O_c \right) 
O_A
\left(I_2 \otimes D_s \otimes I_N\right),
\label{eq:uafact}
\end{equation}
block encodes $A/s$.
\rev{
Here $D_s$ is called a diffusion operator and is defined as 
\begin{equation}
D_s \equiv \underbrace{H\otimes H \otimes \cdots \otimes H}_{m},
\end{equation}
}
\end{theorem}

\begin{proof}
\rev{Note that applying $D_s$ to $\ket{0^m}$ yields
\[
D_s \ket{0^m} = \frac{1}{\sqrt{s}} \sum_{\ell=0}^{s-1} \ket{\ell},
\]
where \{$\ket{\ell}\}$ forms the computational basis in the Hilbert space defined 
by $m$ qubits.}
Our goal is to show that $\bra{0}\bra{0^{m}}\bra{i} U_A \ket{0}\ket{0^{m}}\ket{j}=A_{ij}/s$.
In order to compute the inner product $\bra{0}\bra{0^{m}}\bra{i} U_A \ket{0}\ket{0^{m}}\ket{j}$, we apply $D_{s},O_A,O_c$ to  $\ket{0}\ket{0^{m}}\ket{j}$ successively as illustrated below
\begin{equation}
\begin{split}
\ket{0}\ket{0^{m}}\ket{j}\xrightarrow{D_s} & \frac{1}{\sqrt{s}}\sum_{\ell\in[s]} \ket{0}\ket{\ell}\ket{j}\\
\xrightarrow{O_A} & \frac{1}{\sqrt{s}}\sum_{\ell\in[s]} \left(A_{c(j,\ell),j}\ket{0}+\sqrt{1-\abs{A_{c(j,\ell),j}}^2}\ket{1}\right)\ket{\ell}\ket{j}\\
\xrightarrow{O_c} & \frac{1}{\sqrt{s}}\sum_{\ell\in[s]} \left(A_{c(j,\ell),j}\ket{0}+\sqrt{1-\abs{A_{c(j,\ell),j}}^2}\ket{1}\right)\ket{\ell}\ket{c(j,\ell)},
\end{split}
\label{eq:applydac}
\end{equation}
\rev{where $[s]$ denotes the set of integers $\{0,1,...,s-1\}$.}
Instead of multiplying the leftmost factor $I_2\otimes D_s \otimes I_N$ 
in \eqref{eq:uafact} to last line of \eqref{eq:applydac}, we apply it 
to  $\ket{0}\ket{0^{m}}\ket{i}$ first to obtain
\begin{equation}
\ket{0}\ket{0^{m}}\ket{i}\xrightarrow{D_s}  \frac{1}{\sqrt{s}}\sum_{\ell'\in[s]} \ket{0}\ket{\ell'}\ket{i}.
\label{eq:dsfinal}
\end{equation}
Finally, taking the inner product between \eqref{eq:applydac} and \eqref{eq:dsfinal} yields
\begin{equation}
\bra{0}\bra{0^{m}}\bra{i} U_A \ket{0}\ket{0^{m}}\ket{j}=\frac{1}{s}\sum_{\ell}A_{c(j,\ell),j} \delta_{i,c(j,\ell)}=\frac{1}{s}A_{ij}.
\end{equation}
\end{proof}

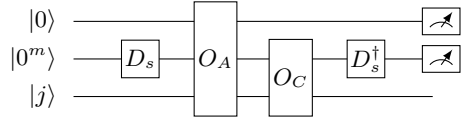
\begin{figure}[!ht]
\centering
\begin{small}
\begin{tikzpicture}[on grid]
\pgfkeys{/myqcircuit, layer width=10mm, row sep=5mm, source node=qwsource}
\newcommand{\qwstart}{1}
\newcommand{\qwend}{6}

\qwire[start node=qw1s, end node=qw1e, style=thin, index=1, start layer=\qwstart, end layer=\qwend, label=$\ket{0}$]
\qwire[start node=qw2s, end node=qw2e, style=thin, index=2, start layer=\qwstart, end layer=\qwend, label=$\ket{0^m}$]
\qwire[start node=qw3s, end node=qw3e, style=thin, index=3, start layer=\qwstart, end layer=\qwend, label=$\ket{j}$]

\singlequbit[style=gate, layer=1, index=2, node=D, label=$D_s$]
\multiqubit[layer=2, start index=1, stop index=3, label=$\,O_A\,$]
\multiqubit[layer=3, start index=2, stop index=3, label=$\,O_C\,$]
\singlequbit[style=gate, layer=4, index=2, node=DT, label=$D_s^{\dagger}$]
\singlequbit[style=meter, layer=5, index=1, node=meas]
\singlequbit[style=meter, layer=5, index=2, node=meas]
\end{tikzpicture}%
\end{small}%
\caption{A general schematic circuit drawing for the block encoding of an $s$-sparse matrix $A$.}
\label{fig:uhcirc2}
\end{figure}

The quantum circuit associated with the block encoding defined in \cref{th:Usparse}
is shown in \cref{fig:uhcirc2}.  Note that the implementation of this 
block encoding requires $m+1$ ancilla qubits 
 in addition to $n$ system qubits.  In order to turn this into an efficient
quantum circuit, we need to \rev{further decompose the $O_c$ and $O_A$ unitaries into a sequence of quantum gates, which may not be straightforward.  %
In the circuit shown in \cref{fig:uhcirc1} which corresponds to the decomposition given in \eqref{eq:uhfac},  $O_A$ is decomposed as $O_A = U_4 U_3 U_2 U_1$ and $O_C$ is simply $U_5$. We will show how these decompositions are constructed systematically in the next section.}

Although it may be possible to construct a set of controlled quantum gates 
to achieve \eqref{eq:oc} and \eqref{eq:oa} for a specific pair of $j$ and $\ell$. 
This approach will not yield an efficient quantum circuit because the total number of quantum gates in the circuit will be on the order of $\mathcal{O}(N)$, which is exponential with respect to $n$. Other brute-force approaches such as the one proposed in~\cite{fable} may also require $\mathcal{O}(N)$ gates. Our goal is to construct a circuit that has a gate complexity of $\mathrm{poly}(n)$, i.e., a polynomial in $n$, at least for certain sparse and/or structured matrices with well defined sparsity patterns.

Note that $O_C$ is unitary. According to \cref{eq:oc}, we must have  $O_C^{\dag}\ket{\ell}\ket{c(j,\ell)}=\ket{\ell}\ket{j}$. This means that for each row index $i=c(j,\ell)$, we can recover the column index $j$ given the value of $\ell$. 
This assumption is of course somewhat restrictive, but it covers important cases such as banded matrices.
For more general query models we refer readers to \cite{GilyenSuLowEtAl2019}. 
We remark that it can generally be more difficult to explicitly 
construct quantum circuits for $O_C$'s in these more general query models.
 
In the following, we give some specific examples to illustrate how efficient 
quantum circuits for $O_C$ and $O_A$ can be constructed for some well structured sparse
matrices.

\subsection{Real symmetric $2 \times 2$ matrix}

\rev{%
We revisit the matrix \eqref{eq:a2} once more and view it as an $s$-sparse matrix with $s = 2$ even though it is dense.  Therefore, we can still use the recipe given in \cref{th:Usparse} to construct a block encoding of $A/2$. We will show below how explicit circuits for $O_C$ and $O_A$ can be constructed.}

\subsubsection{The $O_C$ circuit}

\rev{%
Since the second column of \eqref{eq:a2} is a down or up shift of the first one,}
the function $c(j,\ell)$, which defines the (0-based) row index of the $\ell$th nonzero element in the $j$th column,
\rev{%
can be defined by
\begin{equation}
c(j,\ell) = \mathrm{mod}(j+\ell,2),
\end{equation}
for $j,\ell = \{0,1\}$. 
After enumerating all possible combinations of $(\ell,j)$ input pairs and their corresponding $(\ell,c(j,\ell)$ output pairs shown in \eqref{eq:sym22-OC} (on the right), we see that $O_C$ can be simply implemented as a CNOT gate shown on the left of \eqref{eq:sym22-OC}.
\begin{equation}
\begin{minipage}{0.8\textwidth}
\centering
\begin{small}
\begin{tikzpicture}[on grid]
\pgfkeys{/myqcircuit, layer width=7.5mm, row sep=5mm, source node=qwsource}
\newcommand{\qwstart}{1}
\newcommand{\qwend}{3}
\qwire[index=1, start layer=\qwstart, end layer=\qwend, end node=qw1e, label={$\ket\ell$}]
\qwire[index=2, start layer=\qwstart, end layer=\qwend, end node=qw2e, label={$\ket{j}$}]
\multiqubit[layer=1, start index=1, stop index=2, label=$\,O_C\,$]
\renewcommand{\qwstart}{4}
\renewcommand{\qwend}{5.5}
\qwire[index=1, start layer=\qwstart, end layer=\qwend, start node=qw1s]
\qwire[index=2, start layer=\qwstart, end layer=\qwend, start node=qw2s]
\cnot[layer=3.75, control index=1, target index=2]
\node at ($0.25*(qw1e)+0.25*(qw1s)+0.25*(qw2e)+0.25*(qw2s)$) {$=$};
\node[right] at ($0.5*(qw1e)+0.5*(qw2e)+(35mm,0mm)$) {%
\begin{tabular}{cc}
$\ell$ & $j$ \\ \hline
0 & 0 \\
0 & 1 \\
1 & 0 \\
1 & 1 \\
\end{tabular} $\quad\to\quad$
\begin{tabular}{cc}
$\ell$ & $c(j,\ell)$ \\ \hline
0 & 0 \\
0 & 1 \\
1 & 1 \\
1 & 0 \\
\end{tabular}};
\end{tikzpicture}%
\end{small}
\end{minipage}
\label{eq:sym22-OC}
\end{equation}
using the following two lines \texttt{QCLAB} code}

\begin{minipage}{0.8\textwidth}
\begin{lstlisting}[language=matlab,basicstyle=\ttfamily\small,keywordstyle=\color{blue},commentstyle=\color{green!70!black}]
OC = qclab.QCircuit(2,1);
OC.push_back(qclab.qgates.CNOT(0,1));
\end{lstlisting}
\end{minipage}

\subsubsection{The $O_A$ circuit}

The basic strategy to construct a circuit for $O_A$ defined by \eqref{eq:oa} is
to use controlled rotations to place numerical values at proper locations in $U_A$.
Because the matrix $A$ in \eqref{eq:a2} has at most two unique matrix elements. We need two controlled rotations.
In general, we need to condition the application of these rotations on the values of $j$ and $\ell$.
\rev{%
However, since the values of $\alpha_1$ and $\alpha_2$ in \eqref{eq:a2} depend solely on $\ell$, control is only required on the qubit that takes $\ket{\ell}$ as the input. We apply the rotation $R_y(\theta_1)$ with $\theta_1=2\arccos(\alpha_1)$ when
$\ell = 0$, and $R_y(\theta_2)$ with $\theta_2=2\arccos(\alpha_2)$ when $\ell=1$.
These controlled rotations are combined to yield the following $O_A$ circuit.
\begin{equation}
\begin{minipage}{0.8\textwidth}
\centering
\begin{small}
\begin{tikzpicture}[on grid]
\pgfkeys{/myqcircuit, layer width=7.5mm, row sep=5mm, source node=qwsource}
\newcommand{\qwstart}{1}
\newcommand{\qwend}{3}
\qwire[index=1, start layer=\qwstart, end layer=\qwend, label={$\ket0$}]
\qwire[index=2, start layer=\qwstart, end layer=\qwend, label={$\ket\ell$}, end node=qw2e]
\qwire[index=3, start layer=\qwstart, end layer=\qwend, label={$\ket{j}$}]
\multiqubit[layer=1, start index=1, stop index=3, label=$\,O_A\,$]
\renewcommand{\qwstart}{4}
\renewcommand{\qwend}{8.5}
\qwire[index=1, start layer=\qwstart, end layer=\qwend]
\qwire[index=2, start layer=\qwstart, end layer=\qwend, start node=qw2s]
\qwire[index=3, start layer=\qwstart, end layer=\qwend]
\singlequbit[style=gate, layer=4.25, index=1, node=R1, label=$\,R_y(\theta_1)\,$]
\singlequbit[style=gate, layer=6.25, index=1, node=R2, label=$\,R_y(\theta_2)\,$]
\control[layer=4.25, index=2, target node=R1, style=controloff]
\control[layer=6.25, index=2, target node=R2, style=controlon]
\node at ($0.5*(qw2e)+0.5*(qw2s)$) {$=$};
\node[right] at ($(qw2e)+(55mm,0mm)$) {%
$\begin{aligned}
\theta_1 &= 2\arccos(\alpha_1) \\
\theta_2 &= 2\arccos(\alpha_2) \\
\end{aligned}$};
\end{tikzpicture}%
\end{small}
\end{minipage}
\label{eq:sym22-OA}
\end{equation}
Note that the last qubit is not used in this $O_A$ circuit.
The corresponding \texttt{QCLAB} code is given by}

\begin{minipage}{0.8\textwidth}
\begin{lstlisting}[language=matlab,basicstyle=\ttfamily\small,keywordstyle=\color{blue},commentstyle=\color{green!70!black}]
OA = qclab.QCircuit(3);
OA.push_back(qclab.qgates.CRotationY(1,0,theta1,0));
OA.push_back(qclab.qgates.CRotationY(1,0,theta2,1));
\end{lstlisting}
\end{minipage}

\rev{%
On some quantum hardware, it is preferable to use single qubit rotation and CNOT gates in place of controlled rotation gates. These are sometimes referred to as \emph{uniformly controlled rotations}~\cite{Mottonen2004}.  In \eqref{eq:sym22-ucr}, we show how the controlled rotations used in the $Q_A$ circuit \eqref{eq:sym22-OA} can be replaced by an equivalent set of uniformly controlled rotation gates.}
\begin{equation}
\begin{minipage}{0.85\textwidth}
\centering
\begin{small}
\begin{tikzpicture}[on grid]
\pgfkeys{/myqcircuit, layer width=7.5mm, row sep=5mm, source node=qwsource}
\newcommand{\qwstart}{1}
\newcommand{\qwend}{5.5}
\qwire[index=1, start layer=\qwstart, end layer=\qwend]
\qwire[index=2, start layer=\qwstart, end layer=\qwend, end node=qw2e]
\singlequbit[style=gate, layer=1.25, index=1, node=R1, label=$\,R_y(\theta_1)\,$]
\singlequbit[style=gate, layer=3.25, index=1, node=R2, label=$\,R_y(\theta_2)\,$]
\control[layer=1.25, index=2, target node=R1, style=controloff]
\control[layer=3.25, index=2, target node=R2, style=controlon]
\renewcommand{\qwstart}{6.5}
\renewcommand{\qwend}{11.25}
\qwire[index=1, start layer=\qwstart, end layer=\qwend]
\qwire[index=2, start layer=\qwstart, end layer=\qwend, start node=qw2s]
\singlequbit[style=gate, layer=6.75, index=1, node=Ry1, label=$\,R_y(\phi_1)\,$]
\cnot[layer=7.75, control index=2, target index=1]
\singlequbit[style=gate, layer=8.75, index=1, node=Ry2, label=$\,R_y(\phi_2)\,$]
\cnot[layer=9.75, control index=2, target index=1]
\node at ($0.5*(qw2e)+0.5*(qw2s)+(0,2.5mm)$) {$=$};
\node[right] at ($(qw2e)+(50mm,2.5mm)$) {%
$\begin{aligned}
\theta_1 &= \phi_1 + \phi_2 \\
\theta_2 &= \phi_1 - \phi_2 \\
\end{aligned}$};
\end{tikzpicture}
\end{small}
\end{minipage}
\label{eq:sym22-ucr}
\end{equation}
\rev{The desired uniformly controlled rotations can be constructed in \texttt{QCLAB} by}

\begin{minipage}{0.8\textwidth}
\begin{lstlisting}[language=matlab,basicstyle=\ttfamily\small,keywordstyle=\color{blue},commentstyle=\color{green!70!black}]
OA = qclab.QCircuit(3);
OA.push_back(ucry([theta1,theta2]));
\end{lstlisting}
\end{minipage}

\rev{To see that the right hand side of \eqref{eq:sym22-ucr} is identical to the left hand, we observe first that, if the second qubit is in the $\ket{0}$ state, the circuit on the left applies $R_y(\theta_1)$ to the first qubit. In the circuit on the right, $R_y(\phi_2) R_y(\phi_1)$ is applied to the first qubit in this case. These two operations are identical when $\theta_1 = \phi_1 + \phi_2$. Secondly, if the second qubit is in the $\ket{1}$ state, the left circuit applies $R_y(\theta_2)$ to the first qubit, while the right circuit applies the gate sequence $X R_y(\phi_2) X R_y(\phi_1)$. It follows from the identity $X R_y(\theta) X = R_y(-\theta)$ that these operations are equivalent if $\theta_2 = \phi_1 - \phi_2$.}


\subsubsection{The complete circuit}

\rev{%
Finally, we substitute circuits \cref{eq:sym22-ucr,eq:sym22-OC}, together with $D_s = H$ into \cref{fig:uhcirc2} and obtain the complete block-encoding quantum circuit for the matrix \eqref{eq:a2}.
Note that this circuit is identical to the one we have already introduced in \cref{fig:uhcirc1}.
}
\subsection{Banded circulant matrix}
\label{sec:circulant}
A sparse matrix $A$ can be viewed as an adjacency matrix for a directed graph defined in terms a vertex set $V$, which consists of column indices $\{j\}$ of $A$, and an edge set $E$, which consists of pairs of indices $\{(i,j)\}$.  The $(i,j)$th element of $A$ is nonzero if $(i,j)\in E$.  The numerical value of the $(i,j)$th element of $A$, denoted by $A_{i,j}$, represents the weight of the edge emanating from the vertex $j$ and incident to the vertex $i$ for a directed graph. For an undirected graph, the corresponding adjacency matrix is symmetric, i.e., $A_{i,j} = A_{j,i}$.

In this subsection, we will focus on the sparse matrix associated with a directed cyclic graph.
To illustrate this, we consider the following cyclic graph with $N = 8$ vertices and associated banded circulant adjacency matrix
\begin{align}
\qquad&\begin{small}\begin{tikzpicture}[baseline]
\def\r{1.25}
\draw (0,0) circle(\r);
\foreach \n/\point in {2/(0:\r),1/(45:\r),0/(90:\r),7/(135:\r),6/(180:\r),5/(225:\r),4/(270:\r),3/(315:\r)}
\draw[fill=white] \point circle(1.5ex) node {$\n$};
\end{tikzpicture}\end{small}&
A &= \begin{bmatrix}
\alpha & \gamma & 0      & \cdots & \beta     \\
\beta  & \alpha & \ddots & \ddots &  0        \\
0      & \beta  & \ddots & \gamma    & \vdots \\
\vdots & \ddots & \ddots & \alpha & \gamma    \\
\gamma &     0  & \cdots & \beta  & \alpha    \\
\end{bmatrix}.
\label{eq:matcirc}
\end{align}
Although not shown explicitly in~\eqref{eq:matcirc}, two directed edges emanate from each vertex $j$, $0\leq j \leq N-1$, i.e., $(\mathrm{mod}(j+1,N),j)\in E$ and $(j,\mathrm{mod}(j-1,N)) \in E$.  The weights associated with these edges are $\beta$ and $\gamma$.   We also assign a weight $\alpha$ to each vertex. This weight can also be viewed as the weight associated with a self-loop edge from a vertex $j$ to itself.
The matrix is nearly tridiagonal with the value $\alpha$ on its diagonal, $\gamma$ on its superdiagonal and
$\beta$ on the subdiagonal.  The non-zero elements $A_{N-1,0} = \gamma$ and $A_{0,N-1} = \beta$ reflect the 
cyclic feature of the graph. Each column of the matrix has 3 nonzeros. We use $m=\lceil \log_2 3 \rceil= 2$ ancilla qubits to encode the row indices of the nonzero matrix elements in each column. An additional ancilla qubit is needed to encode the numerical values of the nonzero matrix elements.
\rev{When using ~\cref{th:Usparse} to construct the block encoding circuit, we take $s=2^m=4$ here, i.e., we view \eqref{eq:matcirc} as a $4$-sparse matrix even though each column of the matrix has only 3 nonzero matrix elements.}

\subsubsection{The $O_C$ circuit}
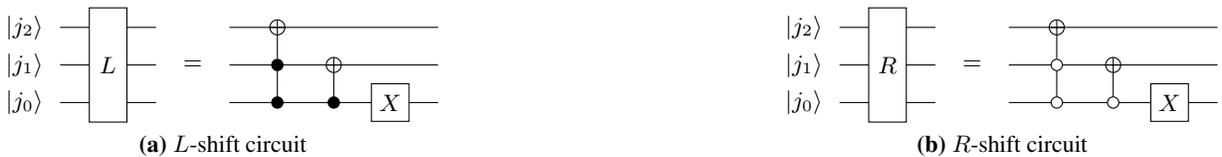
\begin{figure}[hbtp]
\centering
\subfloat[$L$-shift circuit\label{fig:Lshift}]{%
\begin{small}
\begin{tikzpicture}[on grid]
\pgfkeys{/myqcircuit, layer width=7.5mm, row sep=5mm, source node=qwsource}
\newcommand{\qwstart}{1}
\newcommand{\qwend}{3}
\qwire[index=1, start layer=\qwstart, end layer=\qwend, label=\footnotesize{$\ket{j_2}$}]
\qwire[index=2, start layer=\qwstart, end layer=\qwend, label=\footnotesize{$\ket{j_1}$}, end node=qw2e]
\qwire[index=3, start layer=\qwstart, end layer=\qwend, label=\footnotesize{$\ket{j_0}$}]
\multiqubit[layer=1, start index=1, stop index=3, label=$L$]
%
\pgfkeys{/myqcircuit, gate offset=0}
\renewcommand{\qwstart}{4}
\renewcommand{\qwend}{8}
\qwire[index=1, start layer=\qwstart, end layer=\qwend]
\qwire[index=2, start layer=\qwstart, end layer=\qwend, start node=qw2s]
\qwire[index=3, start layer=\qwstart, end layer=\qwend]
\singlequbit[style=not, layer=4, index=1, node=not13]
\control[layer=4, index=2, target node=not13, node=ctrl1, style=controlon]
\control[layer=4, index=3, target node=ctrl1, style=controlon]
\singlequbit[style=not, layer=5, index=2, node=not24]
\control[layer=5, index=3, target node=not24, style=controlon]
\singlequbit[style=gate, layer=6, index=3, label=$X$]
\node at ($0.5*(qw2e)+0.5*(qw2s)$) {$=$};
\end{tikzpicture}%
\end{small}%
}%
\hfill
\subfloat[$R$-shift circuit\label{fig:Rshift}]{%
\begin{small}
\begin{tikzpicture}[on grid]
\pgfkeys{/myqcircuit, layer width=7.5mm, row sep=5mm, source node=qwsource}
\newcommand{\qwstart}{1}
\newcommand{\qwend}{3}
\qwire[index=1, start layer=\qwstart, end layer=\qwend, label=\footnotesize{$\ket{j_2}$}]
\qwire[index=2, start layer=\qwstart, end layer=\qwend, label=\footnotesize{$\ket{j_1}$}, end node=qw2e]
\qwire[index=3, start layer=\qwstart, end layer=\qwend, label=\footnotesize{$\ket{j_0}$}]
\multiqubit[layer=1, start index=1, stop index=3, label=$R$]
%
\pgfkeys{/myqcircuit, gate offset=1}
\renewcommand{\qwstart}{4}
\renewcommand{\qwend}{8}
\qwire[index=1, start layer=\qwstart, end layer=\qwend]
\qwire[index=2, start layer=\qwstart, end layer=\qwend,start node=qw2s]
\qwire[index=3, start layer=\qwstart, end layer=\qwend]
\singlequbit[style=not, layer=3, index=1, node=not13]
\control[layer=3, index=2, target node=not13, node=ctrl1, style=controloff]
\control[layer=3, index=3, target node=ctrl1, style=controloff]
\singlequbit[style=not, layer=4, index=2, node=not24]
\control[layer=4, index=3, target node=not24, style=controloff]
\singlequbit[style=gate, layer=5, index=3, label=$X$]
\node at ($0.5*(qw2e)+0.5*(qw2s)$) {$=$};
\end{tikzpicture}%
\end{small}%
}
\caption{The shift circuits.}
\label{fig:shift}
\end{figure}

For the sparse adjacency matrix induced by the cyclic graph, the function $c(j,\ell)$, which defines
the row index of the $\ell$th nonzero element in the $j$th column, can be written as
\begin{equation}
\revii{c(j,\ell) = 
\begin{cases}
\mathrm{mod}(j-1,N) & \text{if $\ell = 0$ (superdiagonal),} \\
             j      & \text{if $\ell = 1$ (diagonal) or 3,} \\
\mathrm{mod}(j+1,N) & \text{if $\ell = 2$ (subdiagonal).}
\end{cases}}
\label{eq:cjlcyc}
\end{equation}
Therefore, to implement the $O_C$ unitary defined in \eqref{eq:oc}, we need to construct a circuit to 
map $\ket{j}$ to $\ket{\mathrm{mod}(j-1,N)}$\rev{, $\ket{j}$}, or $\ket{\mathrm{mod}(j+1,N)}$ depending on the value of \rev{$\ell = \{0,1,2\}$}.
These cyclic \rev{subtraction} and \rev{addition} mappings are simply \rev{left} and \rev{right} shift permutation operators defined as
\begin{align}
L &=
\begin{bmatrix}
0      & 0      & \hdots & \hdots & 1\\
1      & 0      & 0      & \ddots & 0 \\
\vdots & 1      & \ddots & \ddots & \vdots \\
\vdots & \ddots & \ddots & \ddots & \vdots \\
0      & 0      & \hdots & 1      & 0 \\
\end{bmatrix}, &
R &= 
\begin{bmatrix}
0      & 1      & \hdots & \hdots & 0      \\
0      & 0      & 1      & \ddots & 0      \\
\vdots & \ddots & \ddots & \ddots & \vdots \\
\vdots & \ddots & \ddots & \ddots & 1      \\
1      & 0      & \hdots & \ddots & 0      \\
\end{bmatrix}.
\end{align}
Note that the $L$ and $R$ operators correspond to the addition and subtraction arithmetic operations, i.e., $+1$ and $-1$, respectively. 
They can be implemented by a quantum circuit
consisting of multi-qubit controlled-NOT gates shown in \cref{fig:Rshift,fig:Lshift}.  
The multi-qubit controlled gates perform the carry operation in the binary format (see e.g.~\cite[Chapter 6]{RieffelPolak2011}).

To use these shift components in the quantum circuit implementation of $O_C$, 
we need to control the applications of the $L$ and $R$ shift operators from the 
qubits that take $\ket{\ell}$ as the input.
\rev{For $\ell=0$, we need to apply the $R$-shift
to account for the $-1$ term in \eqref{eq:cjlcyc} and encode the row indices of the superdiagonal elements.
For $\ell=1$, nothing needs to be done.
For $\ell=2$, we need to apply the $L$-shift to account for the $+1$ term and encode the row indeces of the subdiagonal elements.}
Note that 
the use of $D_s$ requires us to also consider $\ell=3$ even though 
the matrix $A$ only has 3 nonzero elements per column. In this case,
\rev{we do not need to do anything}.
All of these conditions can be implemented by the following $O_C$ circuit
\rev{\begin{equation}
\begin{minipage}{0.8\textwidth}
\centering
\begin{small}
\begin{tikzpicture}[on grid]
\pgfkeys{/myqcircuit, layer width=7.5mm, row sep=5mm, source node=qwsource}
\newcommand{\qwstart}{1}
\newcommand{\qwend}{3}
\qwire[index=1, start layer=\qwstart, end layer=\qwend, label=$\rev{\ket{\ell_1}}$]
\qwire[index=2, start layer=\qwstart, end layer=\qwend, label=$\rev{\ket{\ell_0}}$, end node=qw2a]
\qwire[index=3, start layer=\qwstart, end layer=\qwend, label=$\ket{j}$]
\multiqubit[layer=1, start index=1, stop index=3, label=$\,O_C\,$]
\renewcommand{\qwstart}{4}
\renewcommand{\qwend}{7}
\qwire[index=1, start layer=\qwstart, end layer=\qwend]
\qwire[index=2, start layer=\qwstart, end layer=\qwend, start node=qw2b]
\qwire[index=3, start layer=\qwstart, end layer=\qwend]
\singlequbit[style=gate, layer=4, index=3, node=R, label=$R$]
\singlequbit[style=gate, layer=5, index=3, node=L, label=$L$]
\control[layer=4, index=2, target node=R, style=controloff, node=Z]
\control[layer=4, index=1, target node=Z, style=controloff]
\control[layer=5, index=2, target node=L, style=controloff, node=Z]
\control[layer=5, index=1, target node=Z, style=controlon]
\node at ($0.5*(qw2a)+0.5*(qw2b)$) {$=$};
\end{tikzpicture}%
\end{small}
\end{minipage}
\label{eq:oc_cyclic}
\end{equation}
with the corresponding \texttt{QCLAB} code}

\begin{minipage}{0.8\textwidth}
\begin{lstlisting}[language=matlab,basicstyle=\ttfamily\small,keywordstyle=\color{blue},commentstyle=\color{green!70!black}]
OC = qclab.QCircuit(n+2,1);
OC.push_back(rightshift(n+2,2:n+1,[0,1],[0,0]));
OC.push_back( leftshift(n+2,2:n+1,[0,1],[1,0]));
\end{lstlisting}
\end{minipage}

We should note that the construction of $O_C$ depends on how $c(j,\ell)$ is
defined, which is not unique. For example, instead of defining $c(j,\ell)$ by
\eqref{eq:cjlcyc}, we can use the following definition
\begin{equation}
c(j,\ell) =
\begin{cases}
                  j        & \text{if $\ell = 0$ (diagonal) or 3,} \\
\rev{\mathrm{mod}(j+1,N)}  & \text{if $\ell = 1$ (subdiagonal),} \\
\rev{\mathrm{mod}(j-1,N)}  & \text{if $\ell = 2$ (superdiagonal),}
\end{cases}
\label{eq:cjlcyc2}
\end{equation}
resulting in the $O_C$ circuit
\begin{equation}
\begin{minipage}{0.8\textwidth}
\centering
\begin{small}
\begin{tikzpicture}[on grid]
\pgfkeys{/myqcircuit, layer width=7.5mm, row sep=5mm, source node=qwsource}
\newcommand{\qwstart}{1}
\newcommand{\qwend}{3}
\qwire[index=1, start layer=\qwstart, end layer=\qwend, label=$\rev{\ket{\ell_1}}$]
\qwire[index=2, start layer=\qwstart, end layer=\qwend, label=$\rev{\ket{\ell_0}}$, end node=qw2a]
\qwire[index=3, start layer=\qwstart, end layer=\qwend, label=$\ket{j}$]
\multiqubit[layer=1, start index=1, stop index=3, label=$\,O_C\,$]
\renewcommand{\qwstart}{4}
\renewcommand{\qwend}{7}
\qwire[index=1, start layer=\qwstart, end layer=\qwend]
\qwire[index=2, start layer=\qwstart, end layer=\qwend, start node=qw2b, end node=qw2c]
\qwire[index=3, start layer=\qwstart, end layer=\qwend]
\singlequbit[style=gate, layer=4, index=3, node=L, label=$L$]
\singlequbit[style=gate, layer=5, index=3, node=R, label=$R$]
\control[layer=4, index=2, target node=L, style=controlon, node=Z]
\control[layer=4, index=1, target node=Z, style=controloff]
\control[layer=5, index=2, target node=R, style=controloff, node=Z]
\control[layer=5, index=1, target node=Z, style=controlon]
\node at ($0.5*(qw2a)+0.5*(qw2b)$) {$=$};
\renewcommand{\qwstart}{8}
\renewcommand{\qwend}{11}
\qwire[index=1, start layer=\qwstart, end layer=\qwend]
\qwire[index=2, start layer=\qwstart, end layer=\qwend, start node=qw2d]
\qwire[index=3, start layer=\qwstart, end layer=\qwend]
\singlequbit[style=gate, layer=8, index=3, node=L, label=$L$]
\singlequbit[style=gate, layer=9, index=3, node=R, label=$R$]
\control[layer=8, index=2, target node=L]
\control[layer=9, index=1, target node=R]
\node at ($0.5*(qw2c)+0.5*(qw2d)$) {$=$};
\end{tikzpicture}%
\end{small}
\end{minipage}
\label{eq:oc_cyclic2}
\end{equation}
\rev{that can further be simplified because the $L$-shift and $R$-shift cancel each other for $\ell=0$ and $\ell=3$.
The corresponding \texttt{QCLAB} code is given by}

\begin{minipage}{0.8\textwidth}
\begin{lstlisting}[language=matlab,basicstyle=\ttfamily\small,keywordstyle=\color{blue},commentstyle=\color{green!70!black}]
OC = qclab.QCircuit(n+2,1);
OC.push_back( leftshift(n+2,2:n+1,1));
OC.push_back(rightshift(n+2,2:n+1,0));
\end{lstlisting}
\end{minipage}

%
%

\subsubsection{The $O_A$ circuit}
\rev{For circulant matrices, the matrix elements of $A$ depend exclusively on $\ell$. Therefore, when we use controlled rotations to encode the nonzero matrix elements, we place controls only on qubits that take $\ket{\ell}$ as the input. If $c(j,\ell)$ is defined by \eqref{eq:cjlcyc2}, the follow $O_A$ circuit can be constructed to use the rotation $R_y(\theta_0)$ to encode the diagonal entry $\alpha$. The rotation is applied when $\ell=0$. The rotations $R_y(\theta_1)$ and $R_y(\theta_2)$ are used to encode the sub- and sup-diagonal matrix elements $\beta$ and $\gamma$ respectively. They are applied when $\ell=1$ and $\ell=2$ respectively.}
\begin{equation}
\begin{minipage}{0.8\textwidth}
\centering
\begin{small}
\begin{tikzpicture}[on grid]
\pgfkeys{/myqcircuit, layer width=7.5mm, row sep=5mm, source node=qwsource}
\newcommand{\qwstart}{1}
\newcommand{\qwend}{3}
\qwire[index=1, start layer=\qwstart, end layer=\qwend, label=$\ket0$]
\qwire[index=2, start layer=\qwstart, end layer=\qwend, label=$\rev{\ket{\ell_1}}$]
\qwire[index=3, start layer=\qwstart, end layer=\qwend, label=$\rev{\ket{\ell_0}}$, end node=qw2a]
\qwire[index=4, start layer=\qwstart, end layer=\qwend, label=$\ket{j}$]
\multiqubit[layer=1, start index=1, stop index=4, label=$\,O_A\,$]
\renewcommand{\qwstart}{4}
\renewcommand{\qwend}{11}
\qwire[index=1, start layer=\qwstart, end layer=\qwend]
\qwire[index=2, start layer=\qwstart, end layer=\qwend, start node=qw2b]
\qwire[index=3, start layer=\qwstart, end layer=\qwend]
\qwire[index=4, start layer=\qwstart, end layer=\qwend]
\singlequbit[style=gate, layer=4.5, index=1, node=R0, label=$\,R_y(\theta_0)\,$]
\singlequbit[style=gate, layer=6.5, index=1, node=R1, label=$\,R_y(\theta_1)\,$]
\singlequbit[style=gate, layer=8.5, index=1, node=R2, label=$\,R_y(\theta_2)\,$]
\control[layer=4.5, index=2, target node=R0, style=controloff, node=Z]
\control[layer=4.5, index=3, target node=Z,  style=controloff]
\control[layer=6.5, index=2, target node=R1, style=controloff, node=Z]
\control[layer=6.5, index=3, target node=Z,  style=controlon]
\control[layer=8.5, index=2, target node=R2, style=controlon,  node=Z]
\control[layer=8.5, index=3, target node=Z,  style=controloff]
\node at ($0.5*(qw2a)+0.5*(qw2b)$) {$=$};
\end{tikzpicture}%
\end{small}
\end{minipage}
\label{eq:oa_cyclic}
\end{equation}
\rev{The unitary matrix implemented by the circuit can be written as
\begin{equation}
O_A =: O_A\p2 O_A\p1 O_A\p0,
\label{eq:oa_cyclic_matrix}
\end{equation}
where
\begin{align}
O_A\p0 &= R_y(\theta_0) \otimes (E_0 \otimes E_0) \otimes I_n +
              I_2 \otimes (I_4 - E_0 \otimes E_0) \otimes I_n, \label{eq:OA0} \\
O_A\p1 &= R_y(\theta_1) \otimes (E_0 \otimes E_1) \otimes I_n +
              I_2 \otimes (I_4 - E_0 \otimes E_1) \otimes I_n, \label{eq:OA1} \\
O_A\p2 &= R_y(\theta_2) \otimes (E_1 \otimes E_0) \otimes I_n +
              I_2 \otimes (I_4 - E_1 \otimes E_0) \otimes I_n. \label{eq:OA2}
\end{align}
}

\rev{
To determine the value of the rotation angle $\theta_0$ from $\alpha$, we first compute the quantum state after applying the diffusion operator $D_s = H \otimes H$, the $O_A\p0$ circuit, and the $O_C$ circuit to the initial state $\ket0\ket{00}\ket{j}$, i.e.,}
\begin{align}
\Big( I_2 \otimes O_C \Big) & O_A\p0
\Big( I_2 \otimes D_s \otimes I_n \Big) \ket0 \ket{00} \ket{j} \nonumber\\
 &= \rev{\frac12} \Big( I_2 \otimes O_C \Big) O_A\p0 \ket0
    \Big( \ket{00} + \ket{01} + \ket{10} + \ket{11} \Big) \ket{j}, \nonumber\\
 &= \rev{\frac12 \Big( I_2 \otimes O_C \Big)
    \Big( R_y(\theta_0) \ket0\ket{00} + \ket0\ket{01} +
                        \ket0\ket{10} + \ket0\ket{11} \Big) \ket{j},} \nonumber\\
 &= \rev{\frac12
    \Big( R_y(\theta_0) \ket0\ket{00}\ket{j}   + \ket0\ket{01}\ket{j+1} +
                        \ket0\ket{10}\ket{j-1} + \ket0\ket{11}\ket{j} \Big).}
\label{eq:DsOAOC0}
\end{align}
\rev{Next,} taking the inner product of \eqref{eq:DsOAOC0} with $(I_2 \otimes D_s \otimes I_n)\ket0\ket{00}\ket{j}$ results in
\rev{
\begin{equation}
\frac14 \bra{j}\bra{00}\braket{0|R_y(\theta_0)|0}\ket{00}\ket{j} +
\frac14 \bra{j}\bra{11}\braket{0|0}\ket{11}\ket{j} =
\frac14 \left[ \cos\left(\tfrac{\theta_0}{2}\right) + 1 \right],
\end{equation}
where we use the fact that $\braket{0|R_y(\theta_0)|0} = \cos\left(\frac{\theta_0}{2}\right)$.
Finally, to satisfy the equality $\frac14 \left[\cos\left(\tfrac{\theta_0}{2}\right) + 1\right] = \frac{\alpha}{4}$, we obtain $\theta_0 = 2 \arccos(\alpha - 1)$.
Without loss of generality, we have assumed $\alpha$ is real and positive.
The case for $\alpha < 0$ can be handled by first multiplying $A$ by $-1$.}

\rev{To calculate the rotation angle $\theta_1$ from the subdiagonal element $\beta$, we start again from the quantum state after applying $D_s$, $O_A\p1$, and $O_C$ to $(I_2 \otimes D_s \otimes I_n)\ket0\ket{00}\ket{j}$, i.e.,
\begin{align}
\Big( I_2 \otimes O_C \Big) & O_A\p1
\Big( I_2 \otimes D_s \otimes I_n \Big) \ket0 \ket{00} \ket{j} \nonumber\\
 &= \rev{\frac12} \Big( I_2 \otimes O_C \Big) O_A\p1 \ket0
    \Big( \ket{00} + \ket{01} + \ket{10} + \ket{11} \Big) \ket{j}, \nonumber\\
 &= \rev{\frac12 \Big( I_2 \otimes O_C \Big)
    \Big( \ket0\ket{00} + R_y(\theta_1) \ket0\ket{01} +
          \ket0\ket{10} +               \ket0\ket{11} \Big) \ket{j},} \nonumber\\
 &= \rev{\frac12
    \Big( \ket0\ket{00}\ket{j}   +  R_y(\theta_1) \ket0\ket{01}\ket{j+1} +
          \ket0\ket{10}\ket{j-1} +                \ket0\ket{11}\ket{j} \Big).}
\label{eq:DsOAOC1}
\end{align}
Next, taking the inner product of \eqref{eq:DsOAOC1} with $(I_2 \otimes D_s \otimes I_n)\ket0\ket{00}\ket{j+1}$ results in
\begin{equation}
\frac14 \bra{j+1}\bra{01}\braket{0|R_y(\theta_1)|0}\ket{01}\ket{j+1} =
\frac14 \cos\left(\tfrac{\theta_1}{2}\right),
\end{equation}
where we use the fact that $\braket{0|R_y(\theta_1)|0} = \cos\left(\frac{\theta_1}{2}\right)$.
Finally, to satisfy the equality $\frac14 \cos\left(\tfrac{\theta_1}{2}\right) = \frac{\beta}{4}$, we obtain $\theta_1 = 2 \arccos(\beta)$.}

\rev{In a similar way, we can show that $\theta_2 = 2 \arccos(\gamma)$.
 The $O_A$ circuit \eqref{eq:oa_cyclic} corresponding to the $O_C$ circuit \eqref{eq:oc_cyclic2} can be implemented in \texttt{QCLAB} as}

\begin{minipage}{0.95\textwidth}
\begin{lstlisting}[language=matlab,basicstyle=\ttfamily\small,commentstyle=\color{green!70!black}]
theta0 = 2*acos(alpha - 1);
theta1 = 2*acos(beta);
theta2 = 2*acos(gamma);
OA = qclab.QCircuit(n+3);
OA.push_back(qclab.qgates.MCRotationY([1,2],0,[0,0],theta0));
OA.push_back(qclab.qgates.MCRotationY([1,2],0,[0,1],theta1));
OA.push_back(qclab.qgates.MCRotationY([1,2],0,[1,0],theta2));
\end{lstlisting}
\end{minipage}

\noindent
\rev{If we choose to use \eqref{eq:cjlcyc} to define $c(j,\ell)$, we can use controlled rotations by $R_y(\theta_0)$, $R_y(\theta_1)$ and $R_y(\theta_2)$ to encode $\gamma$, $\alpha$ and $\beta$ respectively. It can be shown that corresponding rotations angles are $\theta_0 = 2 \arccos(\gamma)$, $\theta_1 = 2 \arccos(\alpha - 1)$, and $\theta_2 = 2 \arccos(\beta)$. The $Q_A$ circuit can be implemented in 
\texttt{QCLAB} as}

\begin{minipage}{0.95\textwidth}
\begin{lstlisting}[language=matlab,basicstyle=\ttfamily\small,commentstyle=\color{green!70!black}]
theta0 = 2*acos(gamma);
theta1 = 2*acos(alpha - 1);
theta2 = 2*acos(beta);
OA = qclab.QCircuit(n+3);
OA.push_back(qclab.qgates.MCRotationY([1,2],0,[0,0],theta0));
OA.push_back(qclab.qgates.MCRotationY([1,2],0,[0,1],theta1));
OA.push_back(qclab.qgates.MCRotationY([1,2],0,[1,0],theta2));
\end{lstlisting}
\end{minipage}

\rev{In both cases, we can again replace multi-qubit controlled $R_y$ rotations in \eqref{eq:oa_cyclic} with a uniformly controlled rotation. This approach leads to an $O_A$ circuit that has the following structure
\begin{equation}
\begin{minipage}{0.9\textwidth}
\centering
\begin{small}
\begin{tikzpicture}[on grid]
\pgfkeys{/myqcircuit, layer width=7.5mm, row sep=5mm, source node=qwsource}
\newcommand{\qwstart}{1}
\newcommand{\qwend}{3}
\qwire[index=1, start layer=\qwstart, end layer=\qwend, label=$\ket0$]
\qwire[index=2, start layer=\qwstart, end layer=\qwend, label=$\rev{\ket{\ell_1}}$]
\qwire[index=3, start layer=\qwstart, end layer=\qwend, label=$\rev{\ket{\ell_0}}$, end node=qw2a]
\qwire[index=4, start layer=\qwstart, end layer=\qwend, label=$\ket{j}$]
\multiqubit[layer=1, start index=1, stop index=4, label=$\,O_A\,$]
\renewcommand{\qwstart}{4}
\renewcommand{\qwend}{14.25}
\qwire[index=1, start layer=\qwstart, end layer=\qwend]
\qwire[index=2, start layer=\qwstart, end layer=\qwend, start node=qw2b]
\qwire[index=3, start layer=\qwstart, end layer=\qwend]
\qwire[index=4, start layer=\qwstart, end layer=\qwend]
\singlequbit[style=gate, layer=4, index=1, node=R0, label=$\,R_y(\phi_0)\,$]
\singlequbit[style=gate, layer=6.5, index=1, node=R1, label=$\,R_y(\phi_1)\,$]
\singlequbit[style=gate, layer=9, index=1, node=R2, label=$\,R_y(\phi_2)\,$]
\singlequbit[style=gate, layer=11.5, index=1, node=R3, label=$\,R_y(\phi_3)\,$]
\cnot[layer=5.25, control index=3, target index=1]
\cnot[layer=7.75, control index=2, target index=1]
\cnot[layer=10.25, control index=3, target index=1]
\cnot[layer=12.75, control index=2, target index=1]
\node at ($0.5*(qw2a)+0.5*(qw2b)$) {$=$};
\end{tikzpicture}%
\end{small}
\end{minipage}
\label{eq:oa_cyclic_ucr}
\end{equation}
which we implement in \texttt{QCLAB} as follows
}

\begin{minipage}{0.95\textwidth}
\begin{lstlisting}[language=matlab,basicstyle=\ttfamily\small,commentstyle=\color{green!70!black}]
OA = qclab.QCircuit(n+3);
OA.push_back(ucry([theta_0,theta_1,theta_2]));
\end{lstlisting}
\end{minipage}

\noindent%
\rev{The angles $\phi_i$ are computed from $\theta_i$ through a Walsh-Hadamard transformation~\cite{Mottonen2004}.}
\pgfkeys{/myqcircuit, layer width=10mm, row sep=5mm, source node=qwsource}

\subsubsection{The complete circuit}
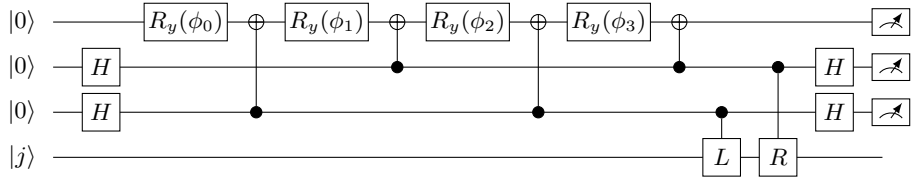
\begin{figure}[hbtp]
\centering%
\begin{small}
\begin{tikzpicture}[on grid]
\pgfkeys{/myqcircuit, layer width=7.5mm, row sep=6mm, source node=qwsource}
\newcommand{\qwstart}{1}
\newcommand{\qwend}{16}

\qwire[start node=qw1s, end node=qw1e, style=thin, index=1, start layer=\qwstart, end layer=\qwend, label=$\ket{0}$]
\qwire[start node=qw2s, end node=qw2e, style=thin, index=2, start layer=\qwstart, end layer=\qwend, label=$\ket{0}$]
\qwire[start node=qw3s, end node=qw3e, style=thin, index=3, start layer=\qwstart, end layer=\qwend, label=$\ket{0}$]
\qwire[start node=qw4s, end node=qw4e, style=thin, index=4, start layer=\qwstart, end layer=\qwend, label=$\ket{j}$]

\singlequbit[style=gate, layer=1,  index=2, node=H1, label=$H$]
\singlequbit[style=gate, layer=1,  index=3, node=H2, label=$H$]
\singlequbit[style=gate, layer=2.5,index=1, node=R1, label=$\,R_y(\phi_0)\,$]
\cnot[layer=3.75, control index=3, target index=1]
\singlequbit[style=gate, layer=5,  index=1, node=R2, label=$\,R_y(\phi_1)\,$]
\cnot[layer=6.25, control index=2, target index=1]
\singlequbit[style=gate, layer=7.5,index=1, node=R3, label=$\,R_y(\phi_2)\,$]
\cnot[layer=8.75, control index=3, target index=1]
\singlequbit[style=gate, layer=10, index=1, node=R4, label=$\,R_y(\phi_3)\,$]
\cnot[layer=11.25, control index=2, target index=1]
\singlequbit[style=gate, layer=12, index=4, node=L, label=$L$]
\singlequbit[style=gate, layer=13, index=4, node=R, label=$R$]
\singlequbit[style=gate, layer=14, index=2, node=H, label=$H$]
\singlequbit[style=gate, layer=14, index=3, node=H, label=$H$]
\singlequbit[style=meter, layer=15, index=1, node=meas]
\singlequbit[style=meter, layer=15, index=2, node=meas]
\singlequbit[style=meter, layer=15, index=3, node=meas]
\control[layer=12, index=3, target node=L]
\control[layer=13, index=2, target node=R]
\end{tikzpicture}%
\end{small}%
\caption{A complete quantum circuit for the block encoding of a $8\times 8$ banded circulant matrix.}
\label{fig:full_cyclic1}
\end{figure}

The complete circuit for a block encoding of $A/\rev{4}$ for a $8\times 8$ circulant matrix of the form \eqref{eq:matcirc} is given in \cref{fig:full_cyclic1}.
The total number of Hadamard gates and controlled rotations are proportional to $\log(s)$ which is generally very small for a sparse matrix. The number of controlled $R$ and $L$ shifts is also on the order of $O(\log s)$. Each controlled-$R$ and controlled-$L$ circuit is a general multi-qubit controlled (Toffoli) gate that can be further decomposed into $\mathrm{poly}(n)$ two-qubit gates. Therefore, the overall gate complexity of the $U_A$ circuit is $\mathrm{poly}(n)$, which is considered efficient.

\begin{figure}[!ht]
\centering%
\begin{small}
\begin{tikzpicture}[on grid]
\pgfkeys{/myqcircuit, layer width=7.5mm, row sep=5mm, source node=qwsource}
\newcommand{\qwstart}{1}
\newcommand{\qwend}{8}
\qwire[index=1, start layer=\qwstart, end layer=\qwend]
\qwire[index=2, start layer=\qwstart, end layer=\qwend, label=$\ket{\ell_1}$]
\qwire[index=3, start layer=\qwstart, end layer=\qwend, label=$\ket{\ell_0}$]
\qwire[index=4, start layer=\qwstart, end layer=\qwend, label=$\ket{j_2}$]
\qwire[index=5, start layer=\qwstart, end layer=\qwend, label=$\ket{j_1}$]
\qwire[index=6, start layer=\qwstart, end layer=\qwend, label=$\ket{j_0}$]

\singlequbit[style=gate, layer=2, index=1, node=R5, label=$\,R_y(\pi - \theta_1)$\,]
\singlequbit[style=gate, layer=5, index=1, node=R6, label=$\,R_y(\pi - \theta_2)$\,]

\control[layer=2, index=2, target node=R5,    style=controloff, node=ctrl1]
\control[layer=2, index=3, target node=ctrl1, style=controlon,  node=ctrl2]
\control[layer=2, index=4, target node=ctrl2, style=controlon,  node=ctrl3]
\control[layer=2, index=5, target node=ctrl3, style=controlon,  node=ctrl4]
\control[layer=2, index=6, target node=ctrl4, style=controlon]

\control[layer=5, index=2, target node=R6,    style=controlon,  node=ctrl1]
\control[layer=5, index=3, target node=ctrl1, style=controloff, node=ctrl2]
\control[layer=5, index=4, target node=ctrl2, style=controloff, node=ctrl3]
\control[layer=5, index=5, target node=ctrl3, style=controloff, node=ctrl4]
\control[layer=5, index=6, target node=ctrl4, style=controloff]
\end{tikzpicture}%
\end{small}%
\caption{\rev{Multi-qubit controlled rotations that can be added to the quantum circuit shown in \cref{fig:full_cyclic1} between $O_A$ and $O_C$ to block encode a tri-diagonal matrix.}}
\label{fig:oa_tridiag}
\end{figure}
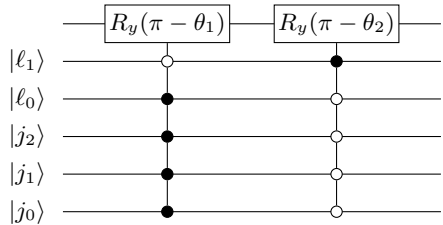

We should also note that the circuit shown in \cref{fig:full_cyclic1} can be
easily modified to block encode a tridiagonal matrix with $\alpha$ on the diagonal, $\beta$ on the subdiagonal
and $\gamma$ on the superdiagonal.  All we need to do is to add the following two multi-qubit control gates
to the sequence of gates in the $O_A$ block of the circuit.  These gates perform controlled rotations \rev{$R_y(\pi - \theta_1)$}
and \rev{$R_y(\pi - \theta_2)$} to zero out $\beta$ in the $(0,N-1)$th entry of $U_A$ and $\gamma$ in the $(N-1,0)$th entry of $U_A$, respectively, \rev{as shown in Figure~\ref{fig:oa_tridiag}.}

\subsection{Extended binary tree}
\label{sec:ebtree}
The banded circulant matrix is special in the sense that the $c(j,\ell)$ function is relatively simple, and the values of nonzero matrix elements are independent of $j$.  Next we consider another special case that has a slightly more complicated
$c(j,\ell)$ function. In this case, the matrix $A$ is the adjacency matrix
for an undirected and balanced binary tree.  A standard balanced binary tree has $2^n-1$ vertices. To simplify our discussion, we add one additional vertex and connect it to the root of the binary tree so that the total number of vertices is $2^n$. The additional vertex is the new root of the tree and labeled by $0$.  We call this binary tree an extended binary tree and consider the following extended binary tree with $8$ vertices and associated $8 \times 8$ adjacency matrix \revii{(i.e., $n=3$)}.
\begin{align}
&\begin{small}\vcenter{\hbox{\begin{forest}
baseline,for tree={
    grow=south,
    circle, draw, minimum size=3ex, inner sep=1pt,
    s sep=5mm
        }
[0
  [1
    [2
        [4]
        [5]
    ]
    [3
        [6]
        [7]
    ]
  ]
]
\end{forest}}}\end{small}&
A &= 
\begin{bmatrix}
\gamma & \beta  &        &        &        &        &        &       \\  
\beta  & \alpha & \beta  & \beta  &        &        &        &       \\
       & \beta  & \alpha &        & \beta  & \beta  &        &       \\
       & \beta  &        & \alpha &        &        & \beta  & \beta \\
       &        & \beta  &        & \gamma &        &        &       \\
       &        & \beta  &        &        & \gamma &        &       \\
       &        &        & \beta  &        &        & \gamma &       \\
       &        &        & \beta  &        &        &        & \gamma 
\end{bmatrix}.
\label{eq:ebtreemat}
\end{align}

We assign the weight $0<\alpha<1$ to each vertex (or self loop from a vertex to itself) with the exception of the root and leaves of the tree. For those vertices, we assign the weight $0<\gamma<1$ instead. We assign the weight $0<\beta<1$ to each edge between a parent and its child.

\subsubsection{The $O_C$ circuit}
\revii{Because each column of the matrix $A$ in \eqref{eq:ebtreemat} has at most 4 nonzero elements, we in principle only need 2 qubits to encode $\ell$ in the function $c(j,\ell)$. However, for this problem, it is convenient to extend the notion of the $\ell$th nonzero element in each column to the $\ell$th type of nonzero element. Note that not all types of nonzero element appear in each column.}

\revii{With this interpretation of $\ell$, we need an additional qubit to encode $\ell$ and the} function $c(j,\ell)$ associated with the nonzero pattern of the adjacency matrix \eqref{eq:ebtreemat} can be defined as
\begin{equation}
c(j,\ell) =
\rev{\begin{cases}
   2j   & \text{if $\ell = 0$ \revii{and $j < 2^{n-1}$} (left child),} \\
   2j+1 & \text{if $\ell = 1$ \revii{and $j < 2^{n-1}$} (right child),} \\
    j/2 & \text{if $\ell = 2$ and $j$ even (parent),} \\
(j-1)/2 & \text{if $\ell = 3$ and $j$ odd (parent),} \\
    j   & \text{if \revii{$3< \ell < 8$} (diagonal),}
\end{cases}}
\label{eq:cjltree}
\end{equation}
\revii{where $j$ and $\ell$ have bit representations $[\ttj_{n-1} \cdots \ttj_1 \ttj_0]$ and $[\ttl_2 \ttl_1 \ttl_0]$, respectively.}
Because this function involves multiplying a column
index $j$ by 2 and dividing $j$ by 2, the $O_C$ circuit contains controlled subcircuits
that perform these operations in addition to the controlled $L$-shift \rev{and $R$-shift} circuit\rev{s} used to perform the addition \rev{and subtraction} by 1 for \rev{$\ell = 1$ and $\ell = 3$, respectively}.

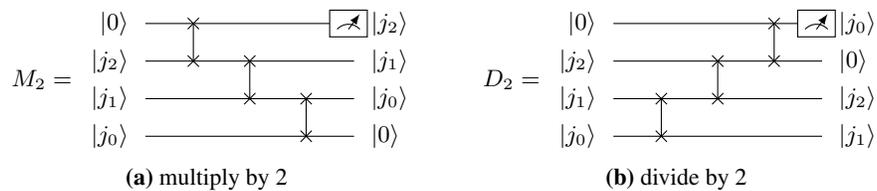
\begin{figure}[hbtp]
\centering%
\begingroup
\captionsetup[subfigure]{width=0.5\linewidth}
\subfloat[multiply by 2\label{fig:mult2}]{%
\begin{small}
\begin{tikzpicture}[on grid]
\pgfkeys{/myqcircuit, layer width=7.5mm, row sep=5mm, source node=qwsource}
\newcommand{\qwstart}{1}
\newcommand{\qwend}{5}

\qwire[start node=qw1s, end node=qw1e, style=thin, index=1, start layer=\qwstart, end layer=\qwend, label=$\ket{0}$]
\qwire[start node=qw2s, end node=qw2e, style=thin, index=2, start layer=\qwstart, end layer=\qwend, label=$\ket{j_2}$]
\qwire[start node=qw3s, end node=qw3e, style=thin, index=3, start layer=\qwstart, end layer=\qwend, label=$\ket{j_1}$]
\qwire[start node=qw4s, end node=qw4e, style=thin, index=4, start layer=\qwstart, end layer=\qwend, label=$\ket{j_0}$]
\node[left=0.75] at ($0.5*(qw2s)+0.5*(qw3s)$) [left] {$M_2 =$};

\swapgate[layer=1, first index=1, second index=2]
\swapgate[layer=2, first index=2, second index=3]
\swapgate[layer=3, first index=3, second index=4]
\singlequbit[style=meter, layer=3.75, index=1, node=meas]
\node[right] at (qw1e) {$\ket{j_2}$};
\node[right] at (qw2e) {$\ket{j_1}$};
\node[right] at (qw3e) {$\ket{j_0}$};
\node[right] at (qw4e) {$\ket{0}$};
\end{tikzpicture}%
\end{small}}%
\endgroup
\quad \quad
\begingroup
\captionsetup[subfigure]{width=0.5\linewidth}
\subfloat[divide by 2\label{fig:div2}]{%
\begin{small}
\begin{tikzpicture}[on grid]
\pgfkeys{/myqcircuit, layer width=7.5mm, row sep=5mm, source node=qwsource}
\newcommand{\qwstart}{1}
\newcommand{\qwend}{5}

\qwire[start node=qw1s, end node=qw1e, style=thin, index=1, start layer=\qwstart, end layer=\qwend, label=$\ket{0}$]
\qwire[start node=qw2s, end node=qw2e, style=thin, index=2, start layer=\qwstart, end layer=\qwend, label=$\ket{j_2}$]
\qwire[start node=qw3s, end node=qw3e, style=thin, index=3, start layer=\qwstart, end layer=\qwend, label=$\ket{j_1}$]
\qwire[start node=qw4s, end node=qw4e, style=thin, index=4, start layer=\qwstart, end layer=\qwend, label=$\ket{j_0}$]
\node[left=0.75] at ($0.5*(qw2s)+0.5*(qw3s)$) [left] {$D_2 =$};

\swapgate[layer=1, first index=3, second index=4]
\swapgate[layer=2, first index=2, second index=3]
\swapgate[layer=3, first index=1, second index=2]
\singlequbit[style=meter, layer=3.75, index=1, node=meas]
\node[right] at (qw1e) {$\ket{j_0}$};
\node[right] at (qw2e) {$\ket{0}$};
\node[right] at (qw3e) {$\ket{j_2}$};
\node[right] at (qw4e) {$\ket{j_1}$};
\end{tikzpicture}%
\end{small}}%
\endgroup
\caption{The quantum circuits for (a) multiplying $\ket{j}$ by 2 and (b) dividing $\ket{j}$ by 2 \revii{and their input and output. Note that the top ancilla qubit is to be measured at the end of the circuit.}}
\label{fig:mult2-div2}
\end{figure}

The mapping from $\ket{j}$ to $\ket{2j}$ can be achieved by a unitary operator $M_2$ which 
consists of a sequence of swap operations between adjacent qubits as shown in \cref{fig:mult2} \revii{for the case of $n = 3$ bit integers}.
The top qubit, \revii{initialized in} $\ket{0}$, is an ancilla qubit. 
\revii{When that qubit is turned into $\ket{1}$, for example, as a result of applying $M_2$ to $\ket{j}$ for $j\geq 2^{n-1}$, the inner product of the $O_C$ output with $\ket{0}\ket{i}$, which is effectively what is carried out in the measurement process, yields 0 for any $i$. Hence, the result of the $M_2$ operation is simply discarded post measurement for $j\geq 2^{n-1}$. Therefore, the ancilla qubit allows us to encode the fact that leaf nodes do not have any child.}


The mapping from $\ket{j}$ to $\ket{j/2}$, which we denote by $D_2$, is the inverse operation of $M_2$. Thus, we can simply reverse the sequence of SWAP gates in \cref{fig:mult2} to obtain a circuit for $D_2$ \rev{in \cref{fig:div2}}. Note that in this case,
\rev{the least significant $\ket{j_0}$ ends up at the \revii{ancilla} qubit}
after $D_2$ is applied.  
\revii{When $j_0=1$, which indicates that $j$ is odd, the ancilla qubit is in state $\ket{1}$ after $D_2$ is applied. Again, the inner product of the $O_C$  output with $\ket{0}\ket{i}$ is 0 for any $i$, which suggests that the division by 2 is completely discarded post measurement. As a result, $D_2$ effectively maps $j$ to $j/2$ only when $j$ is even.}

\revii{Using the} $M_2$ and $D_2$ circuits \revii{in combination with the $L$-shift and $R$- shift circuits}, we can \rev{implement} the $O_C$ circuit \revii{for \eqref{eq:cjltree}} as follows
\revii{\begin{equation}
\begin{minipage}{0.8\textwidth}
\centering
\begin{small}
\begin{tikzpicture}[on grid]
\pgfkeys{/myqcircuit, layer width=7.5mm, row sep=5mm, source node=qwsource}
\newcommand{\qwstart}{1}
\newcommand{\qwend}{3}
\qwire[index=1, start layer=\qwstart, end layer=\qwend, label=$\ket{\ell_2}$]
\qwire[index=2, start layer=\qwstart, end layer=\qwend, label=$\ket{\ell_1}$]
\qwire[index=3, start layer=\qwstart, end layer=\qwend, label=$\ket{\ell_0}$, end node=qw3a]
\qwire[index=4, start layer=\qwstart, end layer=\qwend, label=$\ket{0}$, end node = qwe]
\qwire[index=5, start layer=\qwstart, end layer=\qwend, label=$\ket{j}$]
\multiqubit[layer=1, start index=1, stop index=5, label=$\,O_C\,$]
%
\renewcommand{\qwstart}{5}
\renewcommand{\qwend}{10}
\qwire[index=1, start layer=\qwstart, end layer=\qwend]
\qwire[index=2, start layer=\qwstart, end layer=\qwend]
\qwire[index=3, start layer=\qwstart, end layer=\qwend, start node=qw3b]
\qwire[index=4, start layer=\qwstart, end layer=\qwend, end node=qwe]
\qwire[index=5, start layer=\qwstart, end layer=\qwend]
\multiqubit[layer=5, start index=4, stop index=5, node=M2a, label=$\,M_2\,$]
\singlequbit[style=gate, layer=6, index=5, node=L, label=$L$]
\singlequbit[style=gate, layer=7, index=5, node=R, label=$R$]
\multiqubit[layer=8, start index=4, stop index=5, node=D2b, label=$\,D_2\,$]
\control[layer=5, index=2, target node=M2a,   style=controloff, node=Z]
\control[layer=5, index=1, target node=Z,   style=controloff]
\control[layer=6, index=3, target node=L,   style=controlon,  node=Y]
\control[layer=6, index=2, target node=Y,   style=controloff, node=Z]
\control[layer=6, index=1, target node=Z,   style=controloff]
\control[layer=7, index=3, target node=R,   style=controlon,  node=Y]
\control[layer=7, index=2, target node=Y,   style=controlon,  node=Z]
\control[layer=7, index=1, target node=Z,   style=controloff]
\control[layer=8, index=2, target node=D2b,   style=controlon,  node=Z]
\control[layer=8, index=1, target node=Z,   style=controloff]
\singlequbit[style=meter, layer=9, index=4, node=meas]
\node at ($(qw3a)+(1,0)$) {$=$};
\end{tikzpicture}%
\end{small}
\end{minipage}
\label{eq:ocbtree}
\end{equation}
}
\revii{
This circuit works as follows:
\begin{enumerate}
    \item If $\ell = 0$ or $\ell = 1$, $M_2$ is applied to $\ket{j}$ to effectively yield $\ket{2j}$ if $j < 2^{n-1}$. If $\ell = 1$, $L$ further maps $\ket{2j}$ to $\ket{2j+1}$. (The first two cases in \eqref{eq:cjltree}.)
    \item If $\ell = 3$, $\ket{j}$ is mapped to $\ket{j-1}$ by $R$. This is followed by a division using $D_2$ if $j$ is odd. Otherwise, the result is discarded post measurement.  If $\ell = 2$, $D_2$ is effectively applied to $\ket{j}$  only when $j$ is even. (Cases 3 and 4 in \eqref{eq:cjltree}.)
    \item If $\ell > 3$ ($\ell_2 = 1$), $\ket{j}$ is mapped to itself. (The fifth case in \eqref{eq:cjltree}.)
\end{enumerate}
}

\noindent
The corresponding \texttt{QCLAB} code is

\begin{minipage}{0.8\textwidth}
\begin{lstlisting}[language=matlab,basicstyle=\ttfamily\small,keywordstyle=\color{blue},commentstyle=\color{green!70!black}]
OC = qclab.QCircuit(n+4,1);
OC.push_back(      mul2(n+4,3:n+3,[0,1],[0,0]));
OC.push_back( leftshift(n+4,4:n+3,[0,1,2],[0,0,1]));
OC.push_back(rightshift(n+4,4:n+3,[0,1,2],[0,1,1]));
OC.push_back(      div2(n+4,3:n+3,[0,1],[0,1]));
\end{lstlisting}    
\end{minipage}

\subsubsection{The $O_A$ circuit}
As we indicated earlier, the construction of the $O_A$ quantum circuit depends closely
on how $O_C$ is constructed. For the adjacency matrix associated with the extended binary 
tree, the $O_A$ circuit corresponding to the $O_C$ circuit shown in \cref{eq:ocbtree} 
can be constructed as follows:
\begin{enumerate}
\item First, we use a rotation angle \rev{$\theta_0 = 2\arccos(\beta)$} controlled by 
\rev{$\ell_2 = 0$} \revii{($\ell < 4$)} to place \rev{$\beta$} on the \rev{off-}diagonals of the principal leading block of $U_A$.
\revii{$O_C$ ensures that these are all the edges between parents and children in the tree.}
\item Second, we use a rotation angle \rev{$\theta_1 = 2\arccos(\frac\alpha4)$} controlled by
\rev{$(\ell_2,j_2) = (1,0)$} to place \rev{$\alpha$} at the \rev{diagonal elements not corresponding to the leaves. }\revii{This does over rotate the root by $\theta_1$.}
\item Third, we use a rotation angle \rev{$\theta_2 = 2\arccos(\frac\gamma4)$} controlled by
\rev{$(\ell_2,j_2) = (1,1)$} to place \rev{$\gamma$} at the \rev{diagonal elements corresponding to the leaves.}
\item Finally, we use a rotation angle $\theta_3 = 2\arccos(\frac\gamma4 - \frac\beta2) - \theta_1$ controlled by $(\ell_2,j) = (1,0)$ to \revii{correct the root to $\gamma$, i.e.,} the leading element of $U_A$.
\end{enumerate}
\revii{Note that in step 2 above, the rotation of $\theta_1$ is applied in each of the non-leaf columns for $\ell = 4, 5, 6, 7$. Therefore, to place $\alpha$ on the diagonal of these columns, we need to set the rotation angle to $\theta_1 = 2\arccos(\frac\alpha4)$. Similarly, in step 3, $\gamma$ is divided by 4 in the definition of $\theta_2$.}
The complete $O_A$ circuit resulting from the above procedure can be \rev{implemented by the following circuit
\begin{equation}
\begin{minipage}{0.8\textwidth}
\centering
\begin{small}
\begin{tikzpicture}[on grid]
\pgfkeys{/myqcircuit, layer width=7.5mm, row sep=5mm, source node=qwsource}
\newcommand{\qwstart}{1}
\newcommand{\qwend}{3}
\qwire[index=1, start layer=\qwstart, end layer=\qwend, label=$\ket0$]
\qwire[index=2, start layer=\qwstart, end layer=\qwend, label=$\ket{\ell_2}$]
\qwire[index=3, start layer=\qwstart, end layer=\qwend, label=$\ket{\ell_1}$]
\qwire[index=4, start layer=\qwstart, end layer=\qwend, label=$\ket{\ell_0}$, end node=qw4a]
\qwire[index=5, start layer=\qwstart, end layer=\qwend, label=$\ket{0}$]
\qwire[index=6, start layer=\qwstart, end layer=\qwend, label=$\ket{j_2}$]
\qwire[index=7, start layer=\qwstart, end layer=\qwend, label=$\ket{j_1}$]
\qwire[index=8, start layer=\qwstart, end layer=\qwend, label=$\ket{j_0}$]
\multiqubit[layer=1, start index=1, stop index=8, label=$\,O_A\,$]
\renewcommand{\qwstart}{4}
\renewcommand{\qwend}{13}
\qwire[index=1, start layer=\qwstart, end layer=\qwend]
\qwire[index=2, start layer=\qwstart, end layer=\qwend]
\qwire[index=3, start layer=\qwstart, end layer=\qwend]
\qwire[index=4, start layer=\qwstart, end layer=\qwend]
\qwire[index=5, start layer=\qwstart, end layer=\qwend, start node=qw5b]
\qwire[index=6, start layer=\qwstart, end layer=\qwend]
\qwire[index=7, start layer=\qwstart, end layer=\qwend]
\qwire[index=8, start layer=\qwstart, end layer=\qwend]
\singlequbit[style=gate, layer=4.5, index=1, node=R0, label=$\,R_y(\theta_0)\,$]
\singlequbit[style=gate, layer=6.5, index=1, node=R1, label=$\,R_y(\theta_1)\,$]
\singlequbit[style=gate, layer=8.5, index=1, node=R2, label=$\,R_y(\theta_2)\,$]
\singlequbit[style=gate, layer=10.5, index=1, node=R3, label=$\,R_y(\theta_3)\,$]
\control[layer=4.5, index=2, target node=R0, style=controloff]
\control[layer=6.5, index=2, target node=R1, style=controlon, node=Z]
\control[layer=6.5, index=6, target node=Z,  style=controloff]
\control[layer=8.5, index=2, target node=R2, style=controlon, node=Z]
\control[layer=8.5, index=6, target node=Z,  style=controlon]
\control[layer=10.5,index=2, target node=R3, style=controlon, node=Z]
\control[layer=10.5,index=6, target node=Z,  style=controloff,node=Y]
\control[layer=10.5,index=7, target node=Y,  style=controloff,node=X]
\control[layer=10.5,index=8, target node=X,  style=controloff]
\node at ($0.5*(qw4a)+0.5*(qw5b)$) {$=$};
\end{tikzpicture}%
\end{small}
\end{minipage}
\label{eq:oa_ebtree}
\end{equation}
with the corresponding \texttt{QCLAB} code}

\begin{minipage}{0.95\textwidth}
\begin{lstlisting}[language=matlab,basicstyle=\ttfamily\small]
theta0 = 2*acos(beta);
theta1 = 2*acos(alpha/4);
theta2 = 2*acos(gamma/4);
theta3 = 2*acos(gamma/4 - beta/2) - theta1;
OA = qclab.QCircuit(n+5);
OA.push_back(qclab.qgates.CRotationY(1,0,theta0,0));
OA.push_back(qclab.qgates.MCRotationY([1,5],0,[1,0],theta1));
OA.push_back(qclab.qgates.MCRotationY([1,5],0,[1,1],theta2));
OA.push_back(qclab.qgates.MCRotationY([1,5:n+4],0,...
                                      [1,zeros(1,n)],theta3));
\end{lstlisting}
\end{minipage}

\subsubsection{The complete circuit}
Combining the $O_C$ and $O_A$ circuits, we obtain the complete circuit for the block encoding of the $8\times 8$ adjacency matrix associated with the 8-vertex extended binary tree as shown in \cref{fig:full_ebtree}.

\begin{figure}[!ht]
\centering
\begin{small}
\begin{tikzpicture}[on grid]
\pgfkeys{/myqcircuit, layer width=9mm, row sep=6mm, source node=qwsource}
\newcommand{\qwstart}{1}
\newcommand{\qwend}{12}

\qwire[index=1, start layer=\qwstart, end layer=\qwend, label=$\ket{0}$, end node=qwe1]
\qwire[index=2, start layer=\qwstart, end layer=\qwend, label=$\ket{0}$, end node=qwe2]
\qwire[index=3, start layer=\qwstart, end layer=\qwend, label=$\ket{0}$, end node=qwe3]
\qwire[index=4, start layer=\qwstart, end layer=\qwend, label=$\ket{0}$, end node=qwe4]
\qwire[index=5, start layer=\qwstart, end layer=\qwend, label=$\ket{0}$, end node=qwe5]
\qwire[index=6, start layer=\qwstart, end layer=\qwend, label=$\ket{j_2}$]
\qwire[index=7, start layer=\qwstart, end layer=\qwend, label=$\ket{j_1}$]
\qwire[index=8, start layer=\qwstart, end layer=\qwend, label=$\ket{j_0}$]

\singlequbit[style=gate, layer=1, index=2, node=H1, label=$H$]
\singlequbit[style=gate, layer=1, index=3, node=H2, label=$H$]
\singlequbit[style=gate, layer=1, index=4, node=H3, label=$H$]


\singlequbit[style=gate, layer=2, index=1, node=R0, label=$R_0$]
\singlequbit[style=gate, layer=3, index=1, node=R1, label=$R_1$]
\singlequbit[style=gate, layer=4, index=1, node=R2, label=$R_2$]
\singlequbit[style=gate, layer=5, index=1, node=R3, label=$R_3$]

\control[layer=2, index=2, target node=R0, style=controloff]
\control[layer=3, index=2, target node=R1, style=controlon, node=Z]
\control[layer=3, index=6, target node=Z,  style=controloff]
\control[layer=4, index=2, target node=R2, style=controlon, node=Z]
\control[layer=4, index=6, target node=Z,  style=controlon]
\control[layer=5, index=2, target node=R3, style=controlon, node=Z]
\control[layer=5, index=6, target node=Z,  style=controloff,node=Y]
\control[layer=5, index=7, target node=Y,  style=controloff,node=X]
\control[layer=5, index=8, target node=X,  style=controloff]


\multiqubit[layer= 6, start index=5, stop index=8, node=M1, label=$\,M_2\,$]
\multiqubit[layer= 7, start index=6, stop index=8, node=L,  label=$\,L\,$]
\multiqubit[layer=8, start index=6, stop index=8, node=R,  label=$\,R\,$]
\multiqubit[layer=9, start index=5, stop index=8, node=D2, label=$\,D_2\,$]

\control[layer=6, index=3, target node=M1,  style=controloff, node=Z]
\control[layer=6, index=2, target node=Z,  style=controloff]


\control[layer=7, index=4, target node=L,  style=controlon,  node=Y]
\control[layer=7, index=3, target node=Y,  style=controloff, node=Z]
\control[layer=7, index=2, target node=Z,  style=controloff]


\control[layer=8, index=4, target node=R,  style=controlon,  node=Y]
\control[layer=8, index=3, target node=Y,  style=controlon,  node=Z]
\control[layer=8, index=2, target node=Z,  style=controloff]

\control[layer=9, index=3, target node=D2,  style=controlon,  node=Z]
\control[layer=9, index=2, target node=Z,  style=controloff]

\singlequbit[style=gate, layer=10, index=2, node=H4, label=$H$]
\singlequbit[style=gate, layer=10, index=3, node=H5, label=$H$]
\singlequbit[style=gate, layer=10, index=4, node=H6, label=$H$]

\singlequbit[style=meter, layer=11, index=1, node=meas]
\singlequbit[style=meter, layer=11, index=2, node=meas]
\singlequbit[style=meter, layer=11, index=3, node=meas]
\singlequbit[style=meter, layer=11, index=4, node=meas]
\singlequbit[style=meter, layer=11, index=5, node=meas]


\end{tikzpicture}%
\end{small}%
\caption{\rev{The full block-encoding quantum circuit for the adjacency matrix associated with the extended binary tree shown in \cref{eq:ebtreemat}.}}
\label{fig:full_ebtree}
\end{figure}

This circuit can be easily generalized for larger adjacency matrices associated with an extended binary tree with a larger number ($N=2^n$) of vertices.  However, because the number of
unique cases defined by the function $c(j,\ell)$ in \eqref{eq:cjltree} is a constant $4$ due to the relatively small number of additional vertices each vertex is connected to, the number 
of the controlled multiplication, division and additional gates in $O_C$ is fixed. Furthermore, because the number of distinct values of matrix elements in $A$ is fixed at 3, the number of
controlled rotations required in $O_A$ is also fixed. Each of these controlled rotations can be further decomposed into a subcircuit with $\mathcal{O}(\mathrm{poly}(n))$ two-qubit gates. The overall gate complexity of the quantum circuit is $\mathcal{O}(\mathrm{poly}(n))$, which is considered efficient.


\section{Efficient circuits for the block encoding of a symmetric stochastic matrix and quantum walks}
\label{sec:qw}
In the previous section, we presented a general technique for constructing an efficient quantum circuit to block encode $A/s$ for a properly scaled $s$-sparse matrix $A$.  For many applications such as
solving a linear system, the $1/s$ factor does not fundamentally change the solution besides a simple rescaling (though a large scaling factor can reduce the final success probability). 
\revii{However, 
depending on the context, there may be issues with introducing the $1/s$ factor for certain applications.}
One such an application is the construction of a quantum walk 
from the block encoding of an $s$-sparse symmetric stochastic matrix $P$. For such a 
matrix, we need a different strategy to construct an efficient quantum 
circuit that block encodes $P$ directly instead of $P/s$.

In this section, we will give a brief introduction of the construction 
of a quantum walk from the block encoding of a symmetric $s$-sparse stochatic matrix $P$ associated with a classical random walk. We will explain why it is important to block encode $P$ instead of $P/s$, and present a strategy for 
constructing an efficient quantum circuit for block encoding $P$.  We also
show how such a circuit can be used to construct a quantum circuit for a quantum walk. \revii{We also refer readers to Ref.~\cite{ApersGilyenJeffery2021} for unifying various quantum walk frameworks beyond the scope of this article.}

\rev{A stochastic matrix $P$, also} called a Markov chain matrix, describes the transition probabilities of 
traversing from one vertex to another in a random walk on a graph $G=(V,E)$. 
Since the matrix element $P_{ij}$ represents the probability of walking from 
vertex $i$ to vertex $j$, it satisfies
\begin{equation}
P_{ij}\ge 0, \quad \sum_{j} P_{ij}=1.
\label{eq:pijsum}
\end{equation}

A stochastic matrix has many interesting properties, which are briefly summarized in the supplementary materials. In particular, it can be used to model
a classical random walk, which can be characterized by the equation \revii{$w^T = v^T P^k$}, where the $i$th element of the initial state $v$ specifies the probability of a walker being at vertex $i$ initially, and the $i$th element of the final state $w$ gives the probability of the walker being at vertex $i$ after $k$ steps of random walks have 
been taken using $P$ as the transition probability. The efficiency of the walk
is often measured in terms of the number of steps it takes to reach a certain
vertex (hitting time).

\revii{When $P$ is symmetric, it becomes a doubly stochastic Markov chain matrix. In this section, we consider this particular case and show that a more efficient type of walk called a {\em quantum walk}~\cite{mackay2002quantum,Szegedy2004b,ApersGilyenJeffery2021} can be constructed by block encoding $T_k(P)$, where $T_k(t)=\cos(k\arccos(t))$ is the $k$th degree Chebyshev polynomial of the first kind.}
When such a block encoding is applied to an initial state prepared as $\ket{v} \ket{0}$, it yields
\begin{equation}
w = \left(T_{k}(P)\ket{v}\right)\ket{0} + \ket{\ast}\ket{1}.
\label{eq:Tkv}
\end{equation}
In this context, $\ast$ represents a vector that is irrelevant and can be ignored.
After measuring the second register and obtaining $\ket{0}$, the first register contains $T_k(P)\ket{v}$. This suggests that performing $k$ steps of a quantum walk yields $T_{k}(P)\ket{v}$ instead of $P^k\ket{v}$.  The consequence of such a transformation will be discussed in the supplementary materials through a specific application of a quantum walk.  

Quantum walks have found many other applications in quantum algorithms~\cite{shenvi2003quantum,paparo2012google}.  In this section, we will present the block encoding view of a quantum walk and examine quantum circuits for the block encoding of $T_k(P)$,  which is in turn expressed in terms of the block encoding of $P$.

\subsection{Direct block encoding of $P$ and $T_k(P)$}
\label{sec:directBE}
For an $s$-sparse $P$, we can use the techniques discussed in \cref{sec:circuit} 
to construct an efficient quantum circuit that block encodes $P/s$.  

Once we have the block encoding for $P/s$, we can use the QET discussed in \cref{sec:blkencode} 
to construct a quantum circuit for a block encoding of $T_k(P/s)$. 
However, our ultimate goal for implementing a quantum walk is not to have a block encoding 
of $T_k(P/s)$ but for $T_k(P)$ because $P/s$ is not a stochastic matrix. 
To overcome this difficulty, we may express the $k$th degree Chebyshev polynomial
$T_k(t)$ in terms of $T_j(t/s)$ for $j = 0,1,...,k$, i.e.,
\begin{equation}
T_k(t) = \tilde{T}_k(t') = \sum_{j=0}^k \alpha_j T_j(t'),
\end{equation}
where $t'=t/s$ and $\alpha_j = 0$ for odd (even) $j$ if $k$ is even (odd).

For example, when $s=4$,
\begin{equation}
T_2(t) = \tilde{T}_2(t') = 15 + 16 T_2(t'), \ \ \mbox{where} \ \ t'=t/4.
\label{eq:t2}
\end{equation}

We then try to use the QET to construct a block encoding of $\tilde{T}_k(t')$.
However, to use the QET, the magnitude of the polynomial $\tilde{T}_k(t')$ must be bounded by 1 within the interval $[-1,1]$.
Since the magnitude of $\tilde{T}_k(t')$ can be much larger than 1 at 
$t'=\pm 1$, we need to scale $\tilde{T}_{k}(t')$ by $1/|\tilde{T}_{k}(1)|$ before using the QET. \revii{The subnormalization factor grows rapidly with respect to $s$, which can lead to diminishing successful probability when $s$ is large.  This is the scaling issue we alluded to at the beginning of this section.} For example, for $s=4$, $\tilde{T}_{2}(1)=31$.  As a result, we need to divide $\tilde{T}_{2}(t')$ in \eqref{eq:t2} by 31 before applying the QET. Consequently, we effectively construct a block encoding for $T_k(P)/\alpha$, where $\alpha = \tilde{T}_{k}(1)$, instead of $T_{k}(P)$. This has the effect of lowering the probability of successful measurement of $\ket{0}$ in \eqref{eq:Tkv} (when $T_k(P)$ is replaced with $T_k(P)/\alpha$.)

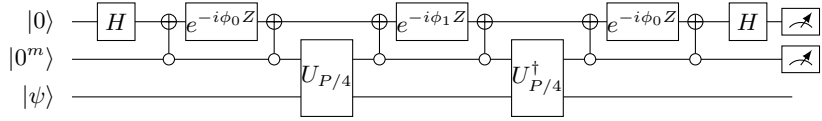
\begin{figure}[!ht]
\centering
\begin{small}
\begin{tikzpicture}[on grid]
\pgfkeys{/myqcircuit, layer width=7mm, row sep=5mm, source node=qwsource}
\newcommand{\qwstart}{1}
\newcommand{\qwend}{15}

\qwire[start node=qw1s, end node=qw1e, style=thin, index=1, start layer=\qwstart, end layer=\qwend, label=$\ket{0}$]
\qwire[start node=qw2s, end node=qw2e, style=thin, index=2, start layer=\qwstart, end layer=\qwend, label=$\ket{0^m}$]
\qwire[start node=qw3s, end node=qw3e, style=thin, index=3, start layer=\qwstart, end layer=\qwend, label=$\ket{\psi}$]

\singlequbit[style=gate, layer=1, index=1, node=H1, label=$H$]
\cnot[layer=2, control index=2, target index=1, style=controloff]
\singlequbit[style=gate, layer=3, index=1, node=R3, label=$e^{-i\phi_0 Z}$]
\cnot[layer=4, control index=2, target index=1, style=controloff]
\multiqubit[layer=5, start index=2, stop index=3, label=$U_{P/4}$]
\cnot[layer=6, control index=2, target index=1, style=controloff]
\singlequbit[style=gate, layer=7, index=1, node=R2, label=$e^{-i\phi_1 Z}$]
\cnot[layer=8, control index=2, target index=1, style=controloff]
\multiqubit[layer=9, start index=2, stop index=3, label=$U^{\dagger}_{P/4}$]
\cnot[layer=10, control index=2, target index=1, style=controloff]
\singlequbit[style=gate, layer=11, index=1, node=R2, label=$e^{-i\phi_0 Z}$]
\cnot[layer=12, control index=2, target index=1, style=controloff]
\singlequbit[style=gate, layer=13, index=1, node=H2, label=$H$]
\singlequbit[style=meter, layer=14, index=1, node=meas]
\singlequbit[style=meter, layer=14, index=2, node=meas]
\end{tikzpicture}%
\end{small}%
\caption{A quantum circuit for $T_2(P)/31$ for a sparse symmetric stochastic matrix $P$ with at most $4$ nonzero matrix elements per column using $U_{P/4}$ which is the block encoding for $P/4$.}
\label{fig:blkenp}
\end{figure}

\revii{We may still} construct a quantum circuit for such a block encoding.
For $k=2$, it follows from the discussion in \cref{sec:blkencode} and \cref{fig:qetcirc} that such a circuit can be drawn as the one shown in
\cref{fig:blkenp}, 
%
where the phase angles 
\[
\phi_0 = 1.17, \: \phi_1 = 0.8 \:, \: \phi_2 = 1.17
\]
can be found numerically by using QSPPACK~\cite{DongMengWhaleyEtAl2021} and $U_{P/s}$ denotes the block encoding of $P/s$.

%

\subsection{An alternative block encoding scheme and its connection to a Szegedy's quantum walk}

There is a way to block encode $P$ instead of $P/s$. Such a block encoding scheme relies on using a different unitary oracle $O_P$ that carries out the following
mapping
\begin{equation}
O_P\ket{0^n}\ket{j}=\sum_{k}\sqrt{P_{jk}}\ket{k}\ket{j}.
\label{eq:op}
\end{equation}
Thanks to the stochasticity of $P$, the right hand side is already a normalized vector, and no additional ancilla qubit is needed \revii{even though the total number of qubits required in this block encoding scheme may exceed that used in \eqref{eq:uafact} for an $s$-sparse matrix with a small $s$. We remark that this approach is not limited to the $s$-sparse case and hence requires two registers $\ket{k}$ and $\ket{j}$ of full size $n$.}

The following \revii{result (see also \cite[Eq. (4)]{ApersGilyenJeffery2021})} describes the structure of the block encoding constructed from a combination of $O_P$ and an $n$-qubit swap operator that performs the following operation
\begin{equation}
\opr{SWAP}\ket{i}\ket{j}=\ket{j}\ket{i},
\label{eq:swap}
\end{equation}
which swaps the value of the two registers in the computational basis, and can be directly implemented using $n$ two-qubit SWAP gates. 

\begin{theorem} \label{th:Up}
If $P$ is a \rev{symmetric} stochastic Markov chain matrix, and if there exists a unitary operator $O_P$ that can be used to carry out the mapping defined in Eq.~\eqref{eq:op}, then  
\begin{equation}
U_P = O_P^{\dagger} \opr{SWAP} O_P
\label{eq:ud}
\end{equation}
is a Hermitian block encoding of $P$, where the swap operator $\opr{SWAP}$ is defined by Eq.~\eqref{eq:swap}.
\end{theorem}
\begin{proof}
Clearly $U_P$ is unitary and Hermitian. 
Now we compute as before
\begin{equation}
\ket{0^{n}}\ket{j}\xrightarrow{O_P} \sum_{k}\sqrt{P_{jk}}\ket{k}\ket{j}\xrightarrow{\opr{SWAP}} \sum_{k}\sqrt{P_{jk}}\ket{j}\ket{k}.
\label{eq:opsource}
\end{equation}
Meanwhile
\begin{equation}
\ket{0^{n}}\ket{i}\xrightarrow{O_P} \sum_{k'}\sqrt{P_{ik'}}\ket{k'}\ket{i}.
\label{eq:optarget}
\end{equation}
The inner product of \eqref{eq:opsource} and \eqref{eq:optarget} yields (using $P_{ij}=P_{ji}$)
\begin{equation}
\bra{0^{n}}\bra{i} U_P \ket{0^n}\ket{j}=\sum_{k,k'}\sqrt{P_{ik'}P_{jk}} \delta_{j,k'}
\delta_{i,k}=\sqrt{P_{ij}P_{ji}}=P_{ij}.
\end{equation}
\end{proof}

Note the \rev{symmetry} requirement for $P$ in the above theorem is not essential.
\revii{When $P$ is not symmetric but is the transition matrix for a reversible Markov chain, we demonstrate in the supplementary materials that we can use the block encoding of a discriminant matrix $D$, which is symmetric. Both $P$ and $D$ have the same set of eigenvalues, and the discriminant matrix $D$ can be utilized for quantum walks.
}
To simplify notation, we assume $P$ \rev{is symmetric} below.
\rev{We should also note that in general the block encoding of a Hermitian matrix $A$ is not necessarily Hermitian.  \cref{th:Up} gives a Hermitian block encoding of a Hermitian $P$. A more general approach of constructing a Hermitian block encoding of a Hermitian matrix $A$ is discussed in the supplementary materials.}
\revii{It is worth noting that the computational advantage of the Hermitian block encoding over a general block encoding of the same Hermitian matrix is still unclear at this stage.}


\subsubsection{\revii{An $O_P$ circuit for a banded circulant stochastic matrix}}
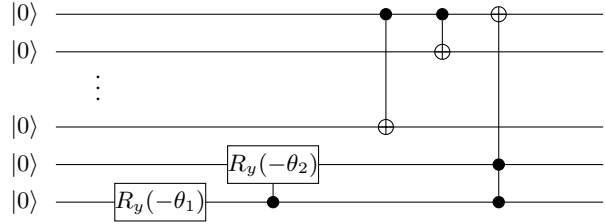
\begin{figure}[htp]
\centering
\begin{small}
\begin{tikzpicture}[on grid]
\pgfkeys{/myqcircuit, layer width=15mm, row sep=5mm, source node=qwsource}
\newcommand{\qwstart}{1}
\newcommand{\qwend}{6}
\qwire[start node=qwsource, end node=qw1e, style=thin,  index=1, start layer=\qwstart, end layer=\qwend, label=$\ket0$]
\qwire[start node=qw2s,       end node=qw2e, style=thin, index=2, start layer=\qwstart, end layer=\qwend,label=$\ket0$]
\drawdots[layer=0, index=3]
\qwire[start node=qw4s,       end node=qw4e, style=thin, index=4, start layer=\qwstart, end layer=\qwend, label=$\ket0$]
\qwire[start node=qw5s,       end node=qw5e, style=thin, index=5, start layer=\qwstart, end layer=\qwend, label=$\ket0$]
\qwire[start node=qw6s,       end node=qw6e, style=thin, index=6, start layer=\qwstart, end layer=\qwend, label=$\ket0$]
\singlequbit[layer=1, index=6, label=$R_y(-\theta_1)$, node=Ry1]

\singlequbit[layer=2, index=5, label=$R_y(-\theta_2)$, node=Ry2]
\control[layer=2, index=6, target node=Ry2, style=controlon]

\singlequbit[style=not, layer=3, index=4, node=targ]
\control[layer=3, index=1, target node=targ, style=controlon, node=c4]

\singlequbit[style=not, layer=3.5, index=2, node=targ2]
\control[layer=3.5, index=1, target node=targ2, style=controlon, node=c6]

\singlequbit[style=not, layer=4, index=1, node=targ3]
\control[layer=4, index=5, target node=targ3, style=controlon, node=c8]
\control[layer=4, index=6, target node=c8, style=controlon]
\end{tikzpicture}%
\end{small}%
\caption{A quantum circuit for the state preparation unitary $K$ given in \eqref{eq:kmat}.}
\label{fig:state_prep}
\end{figure}

We now discuss how to construct an efficient circuit for $O_P$ using the banded circulant matrix \eqref{eq:matcirc}
as an example, where we assume $\alpha + \beta + \gamma = 1$ for the matrix to be stochastic and $\beta = \gamma$.
The construction of a block-encoding circuit can be done in two steps.
In the first step, we construct a unitary $K$ that performs the following mapping
\begin{equation} 
K \ket{0^n} = (\sqrt{\alpha} \: \sqrt{\beta} \: 0 \: \cdots \: 0 \: \sqrt{\gamma})^T.
\label{eq:kmat}
\end{equation}
The first column of $K$ has to encode the square root of the first column of the circulant matrix.
It is generally possible to construct a circuit for performing such a unitary transformation
using \rev{fully-controlled quantum Givens rotations that zero out elements in the vector in Gray
code ordering~\cite{Vartiainen2004}.}  
Our approach is similar, but exploits the sparsity of \eqref{eq:kmat} to \rev{further} reduce the gate complexity to $\bigO(n)$ for this special case.
\rev{The circuit for $K$, shown in \cref{fig:state_prep}, is best explained in reverse
order by considering the action of $K^{\dagger}$.}
\rev{The inverse operation for $K^{\dagger}$ starts with}
a sequence of multi-controlled NOT gates
that permute the non-zero element in the $N-1$st (last) row to row 3.
We can now zero out $\sqrt{\gamma}$ in row 3 by applying the rotation $R_y(\theta_2)$ with
$\theta_2 = -\arctan{\sqrt{\frac{\gamma}{\beta}}}$ to rows 1 and 3 of resulting state. 
Finally, we can zero out the element $\sqrt{\beta + \gamma}$ in row 1 of the resulting state against $\sqrt{\alpha}$ in row 0 with the rotation $R_y(\theta_1)$ where 
$\theta_1 = -\arctan{\sqrt{\frac{\beta + \gamma}{\alpha}}}$.
To obtain a circuit for $K$, we reverse the order
of operations listed above and negate the signs on the rotation angles as shown in \Cref{fig:state_prep}.


\begin{figure}[!ht]
\centering
\begin{small}
\begin{tikzpicture}[on grid]
\pgfkeys{/myqcircuit, layer width=6mm, row sep=5mm, source node=qwsource}
\newcommand{\qwstart}{1}
\newcommand{\qwend}{13}

\qwire[start node=qw1s, end node=qw1e, style=thin, index=1, start layer=\qwstart, end layer=\qwend, label=$\ket{0}$]
\qwire[start node=qw2s, end node=qw2e, style=thin, index=2, start layer=\qwstart, end layer=\qwend, label=$\ket{0}$]
\qwire[start node=qw3s, end node=qw3e, style=thin, index=3, start layer=\qwstart, end layer=\qwend, label=$\ket{0}$]
\qwire[start node=qw4s, end node=qw4e, style=thin, index=4, start layer=\qwstart, end layer=\qwend, label=$\ket{j_2}$]
\qwire[start node=qw5s, end node=qw5e, style=thin, index=5, start layer=\qwstart, end layer=\qwend, label=$\ket{j_1}$]
\qwire[start node=qw6s, end node=qw6e, style=thin, index=6, start layer=\qwstart, end layer=\qwend, label=$\ket{j_0}$]

\multiqubit[layer=1, start index=1, stop index=3, node=K1, label=$K$]
\multiqubit[layer=2, start index=1, stop index=3, node=L3, label=$L$]
\multiqubit[layer=3, start index=1, stop index=2, node=L2, label=$L$]
\control[layer=2, index=6, target node=L3, style=controlon]
\control[layer=3, index=5, target node=L2, style=controlon]
\cnot[layer=4, control index=4, target index=1, style=controlon]
\swapgate[layer=5, first index=3, second index=6]
\swapgate[layer=6, first index=2, second index=5]
\swapgate[layer=7, first index=1, second index=4]
\cnot[layer=8, control index=4, target index=1, style=controlon]
\multiqubit[layer=9, start index=1, stop index=2, node=R2, label=$R$]
\multiqubit[layer=10, start index=1, stop index=3, node=R3, label=$R$]
\control[layer=9, index=5, target node=R2, style=controlon]
\control[layer=10, index=6, target node=R3, style=controlon]
\multiqubit[layer=11, start index=1, stop index=3, node=K2, label=$K^{\dagger}$]
\singlequbit[style=meter, layer=12, index=1, node=meas]
\singlequbit[style=meter, layer=12, index=2, node=meas]
\singlequbit[style=meter, layer=12, index=3, node=meas]
\end{tikzpicture}%
\end{small}%
\caption{\rev{The complete quantum circuit for a Hermitian block encoding of the $8\times 8$ Hermitian banded circulant matrix $P$ defined by \eqref{eq:matcirc} with $\alpha > 0$, $\gamma = \beta > 0$ and $\alpha+2\beta = 1.0$.}}
\label{fig:ua_qwalk}
\end{figure}
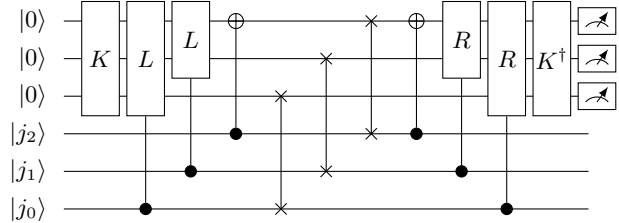

In the second step, we use the fact that $Pe_j = L_n^{j-1} Pe_0$, where $e_j$ is the $j$th column 
of the identity matrix, to construct a controlled shift (adder) circuit to map $\ket{Pe_0}\ket{j}$ to $\ket{Pe_j}\ket{j}$. We show in the supplementary materials how such a controlled $L$-shift circuit is derived. Combining these two steps as well as the SWAP operator, we can represent the quantum circuit for the Hermitian block encoding of $P$ by the diagram shown in \cref{fig:ua_qwalk}.

\begin{figure}[!ht]
\centering
\begin{small}
\begin{tikzpicture}[on grid]
\pgfkeys{/myqcircuit, layer width=6mm, row sep=6mm, source node=qwsource}
\newcommand{\qwstart}{1}
\newcommand{\qwend}{12}

\qwire[start node=qw1s, end node=qw1e, style=thin, index=1, start layer=\qwstart, end layer=\qwend, label=$\ket{0}$]
\qwire[start node=qw2s, end node=qw2e, style=thin, index=2, start layer=\qwstart, end layer=\qwend, label=$\ket{0^n}$]
\qwire[start node=qw3s, end node=qw3e, style=thin, index=3, start layer=\qwstart, end layer=\qwend, label=$\ket{0^n}$]
\singlequbit[style=gate, layer=1, index=1, node=H1, label=$H$]
\cnot[layer=2, control index=2, target index=1, style=controloff]
\singlequbit[style=gate, layer=3, index=1, node=P1, label=$e^{i\frac{\pi}{2} Z}$]
\cnot[layer=4, control index=2, target index=1, style=controloff]
\multiqubit[layer=5, start index=2, stop index=3, node=UD1, label=$U_P$]
\cnot[layer=6, control index=2, target index=1, style=controloff]
\singlequbit[style=gate, layer=7, index=1, node=P2, label=$e^{i\frac{\pi}{2} Z}$]
\cnot[layer=8, control index=2, target index=1, style=controloff]
\multiqubit[layer=9, start index=2, stop index=3, node=UD2, label=$U_P$]
\singlequbit[style=gate, layer=10, index=1, node=H2, label=$H$]
\singlequbit[style=meter, layer=11, index=1, node=meas]
\singlequbit[style=meter, layer=11, index=2, node=meas]
\end{tikzpicture}%
\end{small}%
\caption{The overall structure of a quantum circuit for the block encoding of $T_2(P)$.}
\label{fig:ucheby}
\end{figure}
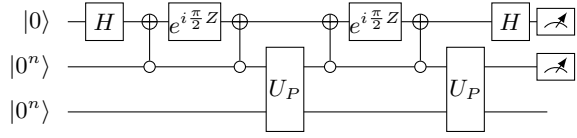

Once a quantum circuit for a symmetric block encoding of $P$ is available, we can then use QET to construct a circuit that block encodes $T_k(P)$ for a $k$th degree of 
Chebyshev polynomial.  It can be easily shown that the phase angles required 
in the QET circuit are
\begin{equation}
\varphi_0 = \pi/2, \: \varphi_1 = \cdots = \varphi_{k-1} = \pi/2 \: 
\mbox{and} \: \varphi_k = 0.
\label{eq:Tvarphi} 
\end{equation}
It follows from the discussion in \cref{sec:blkencode} that the circuit for the block encoding of $T_k(P)$ has a structure shown in \cref{fig:ucheby}.

\revii{In} the following, we draw the connection between the block encoding 
view of a quantum walk with early studies of quantum walks that are often
written as the product of a swap operator and a reflector. Such a quantum 
walk \revii{is} sometimes referred to as the {\em Szegedy quantum walk}~\cite{Szegedy2004b}.  We should note that there are alternative formulations of quantum walks and efficient methods for constructing quantum circuits for these formulations~\cite{Lemieux2020efficientquantum}.

\rev{We begin by noticing that the first two multi-qubit controlled NOT gates and the single qubit phase gate labelled by $e^{i\frac{\pi}{2}Z}$ can be viewed as a way to} 
implement the $e^{i\frac{\pi}{2}Z_{\Pi}}$ operator, where $Z_{\Pi}$ is a reflector defined as
\begin{equation}
Z_{\Pi} \equiv 2 \Pi - I,
\label{eq:zpi}
\end{equation}
with $\Pi \equiv \ket{0^n}\bra{0^n}\otimes I_N$.
In general, such a circuit block is used to implement $e^{i\phi Z_{\Pi}}$, 
which appears in the block encoding of a matrix polynomial,
for an arbitrary angle $\phi$.
However, when $\phi=\pi/2$, we can use the identity 
$e^{i \frac{\pi}{2}Z_{\Pi}} = -i Z_{\Pi}$ to simplify the block encoding circuit.  \rev{This simplification allows us to} rewrite the block encoding of $T_k(P)$ as (up to an irrelevant global  phase factor $(-i)^k$)  
\begin{equation}
U_{T_k(P)} = \left( U_{P} Z_{\Pi} \right)^k.
\label{eq:utkd}
\end{equation}
 




Substituting the symmetric block encoding $U_P$ defined in \eqref{eq:ud} into \eqref{eq:utkd} and grouping $O_P^{\dagger}$, $Z_{\Pi}\otimes I$, $O_P$ and $\opr{SWAP}$ together allows us to express $U_{T_k(P)}$ in terms of repeated applications of 
\begin{equation}
\opr{SWAP} \cdot U_R,
\label{eq:uszegedy}
\end{equation}
where 
$U_R \equiv O_P \left(Z_{\Pi} \otimes I\right) O_P^{\dagger}$ is a reflector. This indicates that the block encoding of $T_k(P)$ is equivalent to performing 
$k$ steps of a Szegedy quantum walk, which is typically defined as the product 
of a reflection and a swap operator. We will show this equivalence 
more precisely in the supplementary materials.

It follows that constructing an efficient circuit for the Szegedy quantum walk is equivalent
to constructing an efficient circuit for the symmetric block encoding of $P$ defined
in \eqref{eq:ud}.
We should note that a method for constructing efficient quantum circuits for
a Szegedy quantum walk on structured graphs was previously presented in
~\cite{Loke2017}. We now show that this method is equivalent to the construction 
of the block encoding circuit for $P$. 

The method presented in~\cite{Loke2017} seeks a unitary transformation $V$ 
in the form of $\sum_{i=0}^{N-1} V_i \otimes \ket{i} \bra{i}$ to diagonalize $U_R$.  
If $V_i$'s can be found to satisfy $V_i \ket{\phi_i} = \ket{b}$ for a computational 
basis $\ket{b}$, then one can verify that
\begin{equation}
V^{\dagger} U_R V = 
2\sum_{i=0}^{N-1}  \Big( V_i^{\dagger} \ket{\phi_i} \bra{\phi_i} V_i \Big) \otimes \ket{i} \bra{i}  = 2(\ket{b}\bra{b} -I).
\label{eq:vuvd}
\end{equation}
If $\ket{b}$ is chosen to be $\ket{0^n}$, the right hand side of \eqref{eq:vuvd} becomes the $Z_{\Pi}$ matrix appeared in \eqref{eq:utkd}.

The transformation $V$ defined in \eqref{eq:vuvd} 
is equivalent to the $O_P$ transformation defined by \eqref{eq:op}.
Hence, constructing a circuit for $O_P$ is equivalent to constructing a circuit for 
$V$. When this circuit is combined with the circuit implementation of $Z_{\Pi}$, 
we obtain a quantum circuit for the Szegedy quantum walk that is equivalent 
to the circuit presented in~\cite{Loke2017}.

\section{Concluding remarks}
\label{sec:conclude}
\rev{Block encodings and quantum eigenvalue/singular value transformations provide a powerful framework for solving large sparse linear algebra problems on quantum computers.  However, to make this approach practical, we need to construct block encoding unitaries that can be easily decomposed into efficient quantum circuits. Although the general strategies for constructing such block encoding unitaries have been proposed in terms of oracles in the past, not much effort has been developed to the explicit construction of quantum circuits for realizing these block encodings. } 

\rev{Our work in this paper is focused on addressing this practical issue. In particular,} we discussed techniques for constructing an efficient quantum circuit
for the block encoding of an $s$-sparse matrix $A/s$ and gave some specific 
examples.  In general, the block encoding circuit consists of an $O_c$ circuit 
block that encodes the nonzero structure of the matrix and an $O_A$ circuit that
encodes the numerical values of the nonzero matrix elements. Without an efficient implementation of these oracles with $\mathrm{poly}(n)$ gate complexity, it can be very difficult to achieve exponential quantum advantage over classical algorithms~\cite{Tang2021,GharibianGall2021}.
Through explicit examples, we show what is required to construct these circuit blocks and 
that the construction of these circuit blocks may not be completely 
independent of each other. In particular, the $O_A$ circuit may be
constructed to zero out certain matrix element to maintain the desired
sparsity structure. 

For a general sparse matrix $A$, constructing an efficient block encoding of $A/s$
can be non-trivial. In particular, it may be difficult to find
an efficient $O_c$ if the sparsity pattern of $A$ is somewhat arbitrary.
In some cases, a more general block encoding scheme that includes an 
additional $O_r$ unitary that encodes the row sparsity separately from the
column sparsity may be needed.

The general construction procedure is not suitable for block encoding 
a sparse symmetric stochastic matrix $P$ associated with a random walk on a graph
when such a block encoding is used to implement a quantum walk on the same graph.  For this type of matrices, an alternative construction procedure that yields 
a symmetric block encoding circuit for $P$ instead of $P/s$ is desired. 
Such a block encoding allows us to construct an efficient block encoding of a 
Chebyshev polynomial of $P$, which is the unitary used in a quantum walk.
We showed that constructing a circuit for a block encoding of $T_k(P)$
is equivalent to a previously developed technique for constructing an efficient 
quantum circuit for Szegedy quantum walks on special graphs. 

Throughout this paper, we assume that single qubit gates are available
for an arbitrary rotation matrix, and that multi-qubit controls are available
to carry out the required encoding schemes.  No constraint has been placed
on the topology of qubits and their connections.  
These assumptions certainly do not hold for existing quantum  devices, and they are not expected to be valid for at least early fault-tolerant quantum computers. When the quantum gates available on these devices are restricted to a few elementary gates (such as Hadamard and Pauli gates), additional decompositions and 
transformations are required to express some of the gates (especially, the multi-qubit control gates) used in the block encoding circuits described above in terms of elementary gates. 
The circuit construction will also need to take the topology and connection
of qubits into consideration. We will pursue efficient ways to construct this type of circuits in future works.

\section{Efficient circuits for powers of a shift operator}
\label{eq:ljcircuit}
In \cref{sec:circulant}, we indicated that powers of a shift operator can be implemented efficiently without repeating the circuit associated with the shift operator. We will elaborate on this point a bit more in this section.

\begin{figure}[htp]
\centering
\begin{small}
\begin{tikzpicture}[on grid]
\pgfkeys{/myqcircuit, layer width=10mm, row sep=5mm, source node=qwsource}
\newcommand{\qwstart}{1}
\newcommand{\qwend}{3}
\qwire[start node=qwsource, end node=qw1e, style=thin,            index=1, start layer=\qwstart, end layer=\qwend, label=$0$]
\qwire[start node=qw2s,       end node=qw2e, style=thin,            index=2, start layer=\qwstart, end layer=\qwend, label=$1$]
\qwire[start node=qw3s,       end node=qw3e, style=thin,            index=3, start layer=\qwstart, end layer=\qwend, label=$2$]
\multiqubit[layer=1, start index=1, stop index=3, label=$L_3^2$]

\pgfkeys{/myqcircuit, gate offset=2}
\renewcommand{\qwstart}{5}
\renewcommand{\qwend}{7}
\qwire[start node=qw1s,       end node=qw1e, style=thin,            index=1, start layer=\qwstart, end layer=\qwend, label=$0$]
\qwire[start node=qw2s,       end node=qw2e, style=thin,            index=2, start layer=\qwstart, end layer=\qwend, label=$1$]
\qwire[start node=qw3s,       end node=qw3e, style=thin,            index=3, start layer=\qwstart, end layer=\qwend, label=$2$]
\multiqubit[layer=3, start index=1, stop index = 2, label=$L_2$]

\node[below right = 1.0*\rowsep and 2.75*\layerwidth of qwsource]{$=$};
\end{tikzpicture}%
\end{small}%
\caption{$L^2$-shift circuit}
\label{fig:L2shift}
\end{figure}
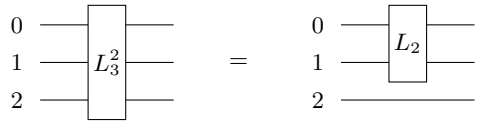

If we use $L_{n}$ to denote an $n$-qubit $L$-shift operator of dimension $2^n\times 2^n$, then it can be shown that 
\begin{equation}
L_n^2 = L_{n-1} \otimes I_2,
\label{eq:l32}
\end{equation}
for $n>1$. As a result, the circuit for the $L_{n}^2$-shift operator can be drawn, for example, as in \cref{fig:L2shift} for $n=3$. A similar decomposition holds for $R_{n}^2$.

Using the binary representation of an integer 
\[
j = [\ttj_{n-1} \cdots \ttj_1 \ttj_0]
  = \ttj_{n-1} \cdot 2^{n-1} + \cdots + \ttj_1 \cdot 2^1 + \ttj_0 \cdot 2^0,
\]
for $j \inN: 0 \leq j \leq 2^n-1$, where $\ttj_{k} \in \{0,1\}$ for $0 \leq k \leq n-1$, we can rewrite $L^j$ as
\begin{equation}
L_n^j = \left(L_n^{2^{n-1}} \right)^{\ttj_{n-1}} \cdot
\left(L_n^{2^{n-2}} \right)^{\ttj_{n-2}} \cdots
\left(L_n^{2^{0}} \right)^{\ttj_0}.
\label{eq:Ljsplit}
\end{equation}

\begin{figure}[htp]
\centering
\begin{small}
\begin{tikzpicture}[on grid]
\pgfkeys{/myqcircuit, layer width=10mm, row sep=5mm, source node=qwsource}
\newcommand{\qwstart}{1}
\newcommand{\qwend}{5}
\qwire[start node=qwsource, end node=qw1e, style=thin,            index=1, start layer=\qwstart, end layer=\qwend, label=$\ket{\ell_2}$]
\qwire[start node=qw2s,       end node=qw2e, style=thin,            index=2, start layer=\qwstart, end layer=\qwend, label=$\ket{\ell_1}$]
\qwire[start node=qw3s,       end node=qw3e, style=thin,            index=3, start layer=\qwstart, end layer=\qwend, label=$\ket{\ell_0}$]
\qwire[start node=qw4s,       end node=qw4e, style=thin,            index=4, start layer=\qwstart, end layer=\qwend, label=$\ket{j_2}$]
\qwire[start node=qw5s,       end node=qw5e, style=thin,            index=5, start layer=\qwstart, end layer=\qwend, label=$\ket{j_1}$]
\qwire[start node=qw6s,       end node=qw6e, style=thin,            index=6, start layer=\qwstart, end layer=\qwend, label=$\ket{j_0}$]
\multiqubit[layer=1, start index=1, stop index=3, label=$L_3$, node=L1]
\control[layer=1, index=6, target node=L1, style=controlon]
\multiqubit[layer=2, start index=1, stop index=2, label=$L_2$, node=L2]
\control[layer=2, index=5, target node=L2, style=controlon]
\cnot[layer=3, control index=4, target index=1, style=controlon]
\end{tikzpicture}%
\end{small}%
\caption{\rev{A quantum circuit for $L_3^j$.}}
\label{fig:L3jshift}
\end{figure}
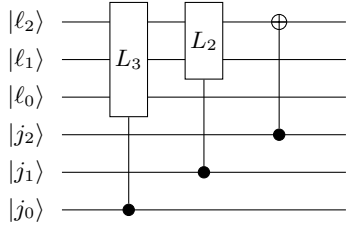

Applying \eqref{eq:l32} recursively to $L_n^{2^{n-k}}$ yields 
\[
L_n^{2^{n-k}} = L_{k}\otimes \underbrace{I_2 \otimes \cdots \otimes I_2}_{n-k}.
\]
Therefore, to apply $L_n^j$ to an input $\ket{\ell}$, we just need to apply a sequence of controlled $L_k$ operations 
for $k = n, n-1,...,1$  successively to the leading $k$ qubits using $\ket{\ttj_k}$ as the control for $L_k$ as shown, for example, in
\cref{fig:L3jshift} for $n=3$.

\section{Hermitian block encoding of Hermitian sparse matrices}
\label{sec:symmetric}
In general, the block encoding unitary of a real symmetric or complex Hermitian matrix
is not necessarily symmetric or Hermitian. This is clearly the case for the block 
encodings constructed in \cref{sec:circulant,sec:ebtree}.
It is also possible to obtain a Hermitian block encoding of a
Hermitian matrix, which can simplify certain theoretical treatments (e.g. quantum walks). We will show how an efficient quantum circuit can be constructed
for such a block encoding. The following theorem lays out the basic structure of such a
circuit.

\begin{theorem}\label{th:symblkenc}
Suppose $A$ is a sparse Hermitian matrix \rev{of dimension $2^n$} with at most $s=2^m$ non-zeros per column, \rev{with $m \leq n$}. If a unitary $O_C$ satisfies
\begin{equation}
O_C \ket{\ell}\ket{j} = \ket{c(j,\ell)}\ket{j},
\label{eq:oc_sym}
\end{equation}
where $c(j,\ell)$ is the row index of the $\ell$th non-zero elements in the $j$th column, and if there exists a unitary $O_A$ such that
\begin{equation}
\left( I\otimes O_A \right) \ket{0}\ket{0}\ket{i}\ket{j} 
= \ket{0}\left(\sqrt{A_{ij}}\ket{0} + \sqrt{1-|A_{ij}|} \ket{1} \right) \ket{i}\ket{j}.
\end{equation}
\rev{For $A_{ij}=\abs{A_{ij}}e^{\I \theta_{ij}},\theta_{ij}\in[0,2\pi)$, the square root is uniquely defined as $\sqrt{A_{ij}}=\sqrt{\abs{A_{ij}}}e^{\I \theta_{ij}/2}$.}
Then the unitary $U_A$ represented by the circuit shown in \cref{fig:ua_sym} is a Hermitian
block encoding of $A$, where $D_s$ is a diffusion operator defined as
\[
\rev{
D_s \equiv I_2\otimes \cdots \otimes \underbrace{H\otimes H \otimes \cdots H}_{m},}
\]
and \opr{SWAP} is a swap operator that swaps the last two $n$-qubit registers in \cref{fig:ua_sym} respectively.
\end{theorem}
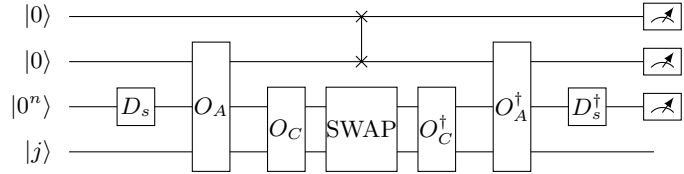
\begin{figure}[!ht]
\centering
\begin{small}
\begin{tikzpicture}[on grid]
\pgfkeys{/myqcircuit, layer width=10mm, row sep=6mm, source node=qwsource}
\newcommand{\qwstart}{1}
\newcommand{\qwend}{9}

\qwire[start node=qw1s, end node=qw1e, style=thin, index=1, start layer=\qwstart, end layer=\qwend, label=$\ket{0}$]
\qwire[start node=qw2s, end node=qw2e, style=thin, index=2, start layer=\qwstart, end layer=\qwend, label=$\ket{0}$]
\qwire[start node=qw3s, end node=qw3e, style=thin, index=3, start layer=\qwstart, end layer=\qwend, label=$\ket{0^n}$]
\qwire[start node=qw3s, end node=qw3e, style=thin, index=4, start layer=\qwstart, end layer=\qwend, label=$\ket{j}$]

\singlequbit[style=gate, layer=1, index=3, node=D, label=$D_s$]
\multiqubit[layer=2, start index=2, stop index=4, label=$O_{A}$]
\multiqubit[layer=3, start index=3, stop index=4, label=$O_C$]
\swapgate[layer=4, first index=1, second index=2]
\multiqubit[layer=4, start index=3, stop index=4, label=$\mathrm{SWAP}$]
\multiqubit[layer=5, start index=3, stop index=4, label=$O_C^{\dagger}$]
\multiqubit[layer=6, start index=2, stop index=4, label=$O_A^{\dagger}$]
\singlequbit[style=gate, layer=7, index=3, node=DT, label=$D_s^{\dagger}$]
\singlequbit[style=meter, layer=8, index=1, node=meas]
\singlequbit[style=meter, layer=8, index=2, node=meas]
\singlequbit[style=meter, layer=8, index=3, node=meas]
\end{tikzpicture}%
\end{small}%
\caption{The general structure of a quantum circuit for a Hermitian block encoding of an $s$-sparse Hermitian matrix $A$. The $S$ block represents a SWAP operator that swaps the the last two $n$-qubit registers qubit-by-qubit.}
\label{fig:ua_sym}
\end{figure}
\begin{proof}
\rev{
Our goal is to show that $\bra{0}\bra{0}\bra{0^{n}}\bra{i} U_A \ket{0}\ket{0}\ket{0^{n}}\ket{j}=A_{ij}/s$.
In order to compute this inner product, we apply $D_{s},O_A,O_c$ to $\ket{0}\ket{0}\ket{0^{n}}\ket{j}$ successively as below
\begin{equation}
\begin{split}
  \ket{0}\ket{0}\ket{0^{n}}\ket{j}\xrightarrow{D_s} & \frac{1}{\sqrt{s}}\sum_{\ell\in[s]} \ket{0}\ket{0}\ket{\ell}\ket{j}\\
  \xrightarrow{O_A} & \frac{1}{\sqrt{s}}\sum_{\ell\in[s]} \ket{0}\left(\sqrt{A_{c(j,\ell),j}}\ket{0}+\sqrt{1-\abs{A_{c(j,\ell),j}}}\ket{1}\right)\ket{\ell}\ket{j}\\
  \xrightarrow{O_c} & \frac{1}{\sqrt{s}}\sum_{\ell\in[s]} \ket{0}\left(\sqrt{A_{c(j,\ell),j}}\ket{0}+\sqrt{1-\abs{A_{c(j,\ell),j}}}\ket{1}\right)\ket{c(j,\ell)}\ket{j}.
\end{split}
\end{equation}
After passing the SWAP gates, the state becomes
\begin{equation}
  \frac{1}{\sqrt{s}}\sum_{\ell\in[s]} \left(\sqrt{A_{c(j,\ell),j}}\ket{0}+\sqrt{1-\abs{A_{c(j,\ell),j}}}\ket{1}\right)\ket{0}\ket{j}\ket{c(j,\ell)}.
\label{eq:applydac_sym}
\end{equation}
We may similarly compute
\begin{equation}
\begin{split}
&\ket{0}\ket{0}\ket{0^n}\ket{0}\ket{i}\xrightarrow{D_s}\xrightarrow{O_A}\xrightarrow{O_c}\\
& \frac{1}{\sqrt{s}}\sum_{\ell'\in[s]} \ket{0}\left(\sqrt{A_{c(i,\ell'),i}}\ket{0}+\sqrt{1-\abs{A_{c(i,\ell'),i}}}\ket{1}\right)\ket{c(i,\ell')}\ket{i}.
\end{split}
\label{eq:dsfinal_sym}
\end{equation}
Finally, taking the inner product between \eqref{eq:applydac_sym} and \eqref{eq:dsfinal_sym} yields
\begin{equation}
\begin{split}
&\bra{0}\bra{0}\bra{0^n}\bra{i}U_A\ket{0}\ket{0}\ket{0^n}\ket{j}\\
=&\frac{1}{s}\sum_{\ell,\ell'\in[s]} \sqrt{A_{c(j,\ell),j}} \sqrt{A^*_{c(i,\ell'),i}}\delta_{i,c(j,\ell)}\delta_{c(i,\ell'),j}\\
=&\frac{1}{s}\sqrt{A_{ij}A_{ji}^*}\sum_{\ell,\ell'} \delta_{i,c(j,\ell)}\delta_{c(i,\ell'),j}=\frac{1}{s}A_{ij}.
\end{split}
\end{equation}
In this equality, we have used the fact that $A$ is Hermitian: $A_{ij}=A_{ji}^*$, and there exists a unique $\ell$ such that $i=c(j,\ell)$, as well as a unique $\ell'$ such that $j=c(i,\ell')$.
%
}
\end{proof}

We use the $8\times 8$ banded circulant matrix defined by \eqref{eq:matcirc} with $0<\alpha,\beta < 1$, $\beta=\gamma$ as an example to illustrate how to construct an efficient quantum circuit for a Hermitian block encoding of a Hermitian sparse matrix.

For this example, we define $c(j,\ell)$ to be used in the construction of the $O_C$ circuit as
\begin{equation}
\rev{
c(j,\ell) = \mathrm{mod}(\ell + j - 1,N),
}
\label{eq:cjlsym}
\end{equation}
to represent the row index of the $\ell$th non-zero matrix elements in the $j$th column. It is important to note that the definition of $O_C$ in \eqref{eq:oc_sym} is different
from the definition given in \eqref{eq:oc}. In \eqref{eq:oc}, the qubit register that takes $\ket{j}$ as the input is changed and the register used to hold $\ket{\ell}$ is simply used as a control and is not altered after the unitary transformation. In \eqref{eq:oc_sym}, the register that holds $\ket{\ell}$ is changed while $\ket{j}$ is unchanged. Therefore, we need to construct a circuit that adds $j-1$ to $\ell$ using $\ket{j}$ as the control.  The addition of $j$ to a quantum state can be carried by applying $L^j$ to that state using the efficient circuit presented in the previous section.

The construction of the $O_A$ circuit in this example is similar to the way $O_A$ is constructed in \cref{sec:circulant}. We use rotations controlled by $\ell$ to place non-zero 
$\sqrt{A_{ij}}$ at appropriate locations in the principal leading block of $U_A$.
To be specific, we use the rotation $R(\theta_1)$ with the angle $\theta_1 = \arccos(\sqrt{\beta}-1)$ conditioned on $\ell=0$ to place $\beta$ on the supdiagonal. We use the rotation $R(\theta_2)$ with $\theta_2 = \arccos(\sqrt{\alpha})+\pi$ conditioned on $\ell=1$ to place $\alpha$ on the diagonal, and the rotation $R(\theta_3)$ with $\theta_3 = \arccos(\beta)+\pi$ to place $\beta$ on the subdiagonal.

\begin{figure}[!ht]
\centering
\begin{small}
\begin{tikzpicture}[on grid]
\pgfkeys{/myqcircuit, layer width=6.4mm, row sep=5mm, source node=qwsource}
\newcommand{\qwstart}{1}
\newcommand{\qwend}{19}

\qwire[start node=qw1s, end node=qw1e, style=thin, index=1, start layer=\qwstart, end layer=\qwend, label=$\ket{0}$]
\qwire[start node=qw2s, end node=qw2e, style=thin, index=2, start layer=\qwstart, end layer=\qwend, label=$\ket{0}$]
\qwire[start node=qw3s, end node=qw3e, style=thin, index=3, start layer=\qwstart, end layer=\qwend, label=$\ket{0}$]
\qwire[start node=qw4s, end node=qw4e, style=thin, index=4, start layer=\qwstart, end layer=\qwend, label=$\ket{0}$]
\qwire[start node=qw5s, end node=qw5e, style=thin, index=5, start layer=\qwstart, end layer=\qwend, label=$\ket{0}$]
\qwire[start node=qw6s, end node=qw6e, style=thin, index=6, start layer=\qwstart, end layer=\qwend, label=$\ket{j_2}$]
\qwire[start node=qw7s, end node=qw7e, style=thin, index=7, start layer=\qwstart, end layer=\qwend, label=$\ket{j_1}$]
\qwire[start node=qw8s, end node=qw8e, style=thin, index=8, start layer=\qwstart, end layer=\qwend, label=$\ket{j_0}$]

\singlequbit[style=gate, layer=1, index=4, node=H1, label=$H$]
\singlequbit[style=gate, layer=1, index=5, node=H2, label=$H$]
\singlequbit[style=gate, layer=2, index=1, node=R1, label=$R_1$]
\singlequbit[style=gate, layer=3, index=1, node=R2, label=$R_2$]
\singlequbit[style=gate, layer=4, index=1, node=R3, label=$R_3$]
\control[layer=2, index=7, target node=R1, style=controloff, node=ctrl1]
\control[layer=2, index=6, target node=ctrl1, style=controloff]
\control[layer=3, index=7, target node=R2, style=controloff, node=ctrl2]
\control[layer=3, index=6, target node=ctrl2, style=controlon]
\control[layer=4, index=7, target node=R3, style=controlon, node=ctrl3]
\control[layer=4, index=6, target node=ctrl3, style=controloff]
\multiqubit[layer=5, start index=3, stop index=5, node=R, label=$R$]
\multiqubit[layer=6, start index=3, stop index=5, node=L3,label=$L$]
\multiqubit[layer=7, start index=3, stop index=4, node=L2,label=$L$]
\cnot[layer=8, control index=6, target index=3, style=controlon]
\control[layer=6, index=8, target node=L3, style=controlon]
\control[layer=7, index=7, target node=L2, style=controlon]
\swapgate[layer=9, first index=1, second index=2]
\multiqubit[layer=9, start index=3, stop index=8, label=$\mathrm{SWAP}$]
\cnot[layer=10, control index=6, target index=3, style=controlon]
\multiqubit[layer=11, start index=3, stop index=4, node=R2,label=$R$]
\multiqubit[layer=12, start index=3, stop index=5, node=R3,label=$R$]
\control[layer=11, index=7, target node=R2, style=controlon]
\control[layer=12, index=8, target node=R3, style=controlon]
\multiqubit[layer=13, start index=3, stop index=5, node=L, label=$L$]
\singlequbit[style=gate, layer=16, index=1, node=R1, label=$R_1^{\dagger}$]
\singlequbit[style=gate, layer=15, index=1, node=R2, label=$R_2^{\dagger}$]
\singlequbit[style=gate, layer=14, index=1, node=R3, label=$R_3^{\dagger}$]
\control[layer=16, index=7, target node=R1, style=controloff, node=ctrl4]
\control[layer=16, index=6, target node=ctrl4, style=controloff]
\control[layer=15, index=7, target node=R2, style=controloff, node=ctrl5]
\control[layer=15, index=6, target node=ctrl5, style=controlon]
\control[layer=14, index=7, target node=R3, style=controlon, node=ctrl6]
\control[layer=14, index=6, target node=ctrl6, style=controloff]
\singlequbit[style=gate, layer=17, index=4, node=H1, label=$H$]
\singlequbit[style=gate, layer=17, index=5, node=H2, label=$H$]
\singlequbit[style=meter, layer=18, index=1, node=meas]
\singlequbit[style=meter, layer=18, index=2, node=meas]
\singlequbit[style=meter, layer=18, index=3, node=meas]
\singlequbit[style=meter, layer=18, index=4, node=meas]
\singlequbit[style=meter, layer=18, index=5, node=meas]
\end{tikzpicture}%
\end{small}%
\caption{\rev{The complete quantum circuit for a Hermitian block encoding of the $8\times 8$ Hermitian banded circulant matrix $A$ defined by \eqref{eq:matcirc}. The SWAP operator swaps the last two $n$-qubit registers qubit-by-qubit.}}
\label{fig:uaband_sym}
\end{figure}
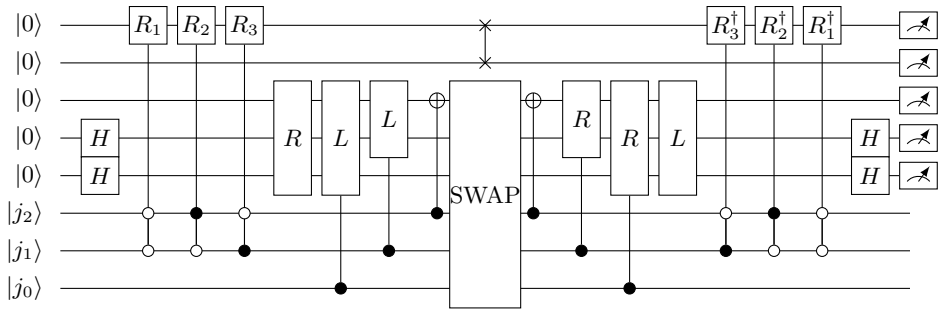

The complete circuit for the Hermitian block encoding of $A$ is shown in \cref{fig:uaband_sym}.

\section{Additional results on quantum walks}

In the following section, we provide additional materials that describe properties of a random walk 
as well as an application that demonstrates the higher (query) efficiency of 
a quantum walk compared to that of a classical random walk from a block encoding point of view.

\subsection{Properties a Markov chain matrix}
Because the sum of all probabilities of transitions from the vertex $i$ to other vertices must be 1 for all $i$, the vector $\left(1 \: 1\: \cdots 1 \right)^T$
is a right eigenvector of the Markov chain stochastic matrix $P$ associated with the eigenvalue $1$.  The corresponding left eigenvector $\pi$ is called a {\em stationary state} and satisifies
\begin{equation}
\revii{\pi^T P=\pi^T}, \quad \pi_i\ge 0, \quad \sum_{i}\pi_i =1.
\label{eq:stationary}
\end{equation}

A Markov chain matrix $P$ is \emph{irreducible} if any state can be reached from any other state in a finite number of steps of a random walk, i.e., \revii{the $j$th element of $w^T=e_i^T P^k$ is nonzero for any $i,j$} and some finite $k$, where $e_i$ is the $i$th column of the identity matrix. An irreducible Markov chain is \emph{aperiodic} if there exists no integer greater than one that divides the length of every directed cycle of the graph. A Markov chain is \emph{ergodic} if it is both irreducible and aperiodic. By the Perron--Frobenius Theorem, any ergodic Markov chain $P$ has a unique stationary state $\pi$ in \eqref{eq:stationary}.
A Markov chain is \emph{reversible} if the following detailed balance condition is satisfied
\begin{equation}
\pi_i P_{ij}=\pi_j P_{ji}.
\end{equation}

For a nonsymmetric but reversible $P$, we can construct a quantum walk by block encoding the \emph{discriminant matrix} $D$ associated with $P$. Such a matrix is defined componentwise as
\begin{equation}
\label{eqn:discriminant_matrix}
D_{ij}=\sqrt{P_{ij}P_{ji}},
\end{equation}
which is real symmetric.
 
When $P$ is reversible, it can be shown that
\begin{equation}
\ket{\pi}=\sum_{i}\sqrt{\pi_i}\ket{i}
\end{equation}
is a normalized eigenvector of the discrimant matrix $D$ defined by \eqref{eqn:discriminant_matrix}, i.e., $\ket{\pi}$ satisfies
\begin{equation}
D\ket{\pi}=\ket{\pi}.
\end{equation}
Furthermore, because $\pi_i>0$ for all $i$, we have
\begin{equation}
D=\diag(\sqrt{\pi}) P \diag(\sqrt{\pi})^{-1}.
\end{equation}
Therefore, for a reversible $P$, $P$ and $D$ share the same set of eigenvalues, and the spectral
properties of a nonsymmetric $P$ can be analyzed by working with the symmetric $D$ matrix.

\subsection{The equivalence of Szegedy quantum walk and block encoding of $T_k(D)$}

The Szegedy quantum walk is traditionally introduced as follows.
Using the following $O_P$ oracle and the multi-qubit $\opr{SWAP}$ gate, we can define two sets of quantum states
\begin{equation}
\begin{split}
\ket{\psi_j^{1}}&=O_P\ket{0^n}\ket{j}=\sum_{k}\sqrt{P_{jk}}\ket{k}\ket{j},\\ \ket{\psi_j^{2}}&=\opr{SWAP}( O_P\ket{0^n}\ket{j})=\sum_{k}\sqrt{P_{jk}}\ket{j}\ket{k}. \end{split}
\end{equation}
These quantum states yield two projection operators
\begin{equation}
\Pi_l=\sum_{j\in[N]}\ket{\psi_j^{l}}\bra{\psi_j^{l}}, \quad l=1,2,
\end{equation}
from which we can define two $2n$-qubit reflection operators $R_{\Pi_l}=2 \Pi_l-I_{2n}$. Let us write down the reflection operators more explicitly. Using the resolution of identity, we obtain
\begin{equation}
R_{\Pi_1}=O_P ((2\ket{0^n}\bra{0^n}-I)\otimes I_n)O_P^{\dag}=O_P (Z_{\Pi}\otimes I_n) O_P^{\dag},
\end{equation}
where $Z_{\Pi}$ is as defined in \eqref{eq:zpi}. 
Similarly, we have
\begin{equation}
R_{\Pi_2}=\opr{SWAP}O_P (Z_{\Pi}\otimes I_n) O_P^{\dag}\opr{SWAP}.
\end{equation}

Then Szegedy's quantum walk operator takes the form
\begin{equation}
\mc{U}_Z=R_{\Pi_2}R_{\Pi_1},
\end{equation}
which is a rotation operator that resembles the one that appears in the Grover's algorithm.
Note that
\begin{equation}
\mc{U}_Z=\opr{SWAP}O_P (Z_{\Pi}\otimes I_n) O_P^{\dag}\opr{SWAP}O_P (Z_{\Pi}\otimes I_n) O_P^{\dag},
\end{equation}
so $O^{\dag}_P \mc{U}_Z (O^{\dag}_P)^{-1}$ is the same as a block encoding of $T_2(D)$ using qubitization up to a matrix similarity transformation. 
More generally, $k$-steps of Szegedy's quantum walk given by $\mc{U}_Z^k$ is equivalent to a block encoding of $T_{2k}(D)$.

\subsection{Quantum walk efficiency in detecting a marked vertex}
In this section, we provide a brief explanation of the advantage of
a quantum walk over a classical random walk from a block encoding point of view, i.e., why it is of
interest to block encode $T_k(D)$ and perform the corresponding 
unitary transformation \eqref{eq:Tkv} on a quantum computer rather than simply 
applying $P^k$ to an initial vector of probability distributions
on a classical computer.

\revii{Consider a symmetric $P$ first.}
The direct consequence of applying $T_k(P)$ in a quantum walk (instead of $P^k$
in a classical random walk) to $v$ can be analyzed as follows.
Since an eigenvalue $\lambda$ of $P$ is between $-1$ and $1$, we can reparameterize it
as $\lambda = \cos \theta$. The corresponding eigenvalue of $P^k$ is $\cos^{k} \theta$, and the corresponding eigenvalue of $T_k(P)$ is $\cos (k\theta)$. For small $\theta$, it follows from the Taylor
expansion of $\cos^k \theta$ and $\cos \hat{k}\theta$ that
\begin{eqnarray}
\cos^k \theta       &=& 1 - \frac{k\theta^2}{2} + O(k^2\theta^4), \\
\cos \hat{k} \theta &=& 1 - \frac{\hat{k}^2\theta^2}{2} + O(\hat{k}^4\theta^4),
\end{eqnarray}
for integers $k$ and $\hat{k}$.

We can readily see that the Taylor expansions of these functions agree up to the second order term when
$\hat{k} = \sqrt{k}$.  This observation suggests that, for a reversible random walk,
the second largest eigenvalue of $T_{\hat{k}}(P)$ can reach that of $P^k$ when 
$\hat{k} = O(\sqrt{k})$. Since properties of a random walk is often determined by the gap between the largest eigenvalue and the second largest eigenvalue, this observation highlights the fundamental reason 
why a quantum walk can be asymptotically much faster than a classical random walk.

To demonstrate how a quantum walk can be used to solve a practical problem, let us examine the 
following example.  Our goal is to detect the presence of a marked
vertex in a graph $G = (V,E)$. Our assumption is that we do not know a priori whether such a 
marked vertex exists. Nor do we know the location of this vertex if it indeed exists.
What we are given are some tools we can use to find out the presence of the marked vertex 
and its position if it is present.

In the classical setting, the tool we are given is a random walk transition probability 
matrix $P$ (constructed by someone who knows the answer). We are allowed to apply $P$ to 
a vector and check the result (but not allowed to look at $P$ itself). The vector we will apply
$P$ to is prepared as $v = e/N$, where $e= (1,\ldots,1)^{T}$. It describes an initial 
uniform probability of being at any vertex.  We examine the probability of being at each vertex 
after performing several steps of the random walk, i.e., we examine elements of the vector \revii{$w^T = v^T P^k$}. If $G$ contains a marked vertex, the element of $w$ associated with such vertex 
will have a much higher magnitude. The position of this element informs us the position 
of the marked vertex.

\rev{To clearly demonstrate the advantage of performing a quantum walk, we choose $G$ to be a complete graph, even though the efficiency of a quantum walk holds for a general graph.} If no marked vertex is present in $G$, the matrix $P$ we are given is
\begin{equation}
P=\frac{1}{N} e e^{T}, \quad e=(1,\ldots,1)^{T}.
\label{eq:pnomark}
\end{equation} 
Otherwise, if a marked vertex is present, and if, without loss of generality, the marked
vertex is the $0$-th vertex (which we do not know in advance), the matrix $P$ we are given
is \revii{nonsymmetric and takes the form}
\begin{equation}
P=\begin{pmatrix}
1 & 0\\
\frac{1}{N} \wt{e} & \frac{1}{N} \wt{e}\wt{e}^{T}
\end{pmatrix},
\label{eq:pmark}
\end{equation}
where $\wt{e}$ is a vector of all ones of length $N-1$.

\revii{Note that $v^{T}P=v$
when $P$ matrix defined by \eqref{eq:pnomark}.}
The output from such a random walk remains the same, i.e., the probability of being at each 
vertex remains at $1/N$. 

However, if we are given the $P$ matrix defined by \eqref{eq:pmark} and is non-symmetric,  \revii{$w^T = v^T P^k$ converges to the left eigenvector} $\wt{\pi}=(1,0,\ldots,0)^{T}$ 
associated with the largest eigenvalue 1 of $P$ when $k = \Or(N)$. Hence, after performing
$\Or(N)$ steps of the random walk using this $P$, we can conclude, with high confidence, that
there is a marked vertex and it is the $0$th vertex since the 0th element of $w$ is nearly 1.

In the quantum setting, the tool we are given is different. Instead of the $P$ matrix, 
we \revii{use} the block encoding of the discriminant matrix $D$ associated with $P$ denoted by 
$U_D$. Again, we are allowed to apply $U_D^k$ to a carefully prepared quantum state $\ket{\psi_0}$ 
to perform a quantum walk (but not allowed to look at $U_D$ itself). The prepared initial 
quantum state is
\begin{equation}
  \ket{\psi_0}=\ket{0^n}(H^{\otimes n} \ket{0^n}).
\label{eq:psi0}
\end{equation}
To detect the presence of a marked vertex, which is equivalent
to determining whether the quantum walk operator is the block encoding of the discriminant
matrix associated with the $P$ matrix defined in \eqref{eq:pnomark} or \eqref{eq:pmark},
we measure the first $n$ qubits, and evaluate the success probability of measuring the $\ket{0^n}$ state. This probability should be
\begin{equation}
p(\ket{0^n})=\norm{(\ket{0^n}\bra{0^n}\otimes I_n)U_D^k\ket{\psi_0}}^2.
\label{eq:prob}
\end{equation}

Then, if $U_D$ were the block encoding of $P$ (which is the same as the corresponding discriminant matrix $D$ when $P=P^T$), $U_D \ket{\psi_0} = \ket{\psi_0}$. As a result, the probability of
measuring $\ket{0^n}$ is $p(\ket{0^n})=1$, regardless how many steps of the quantum walk are taken.

On the other hand, if $U_D$ results from the block encoding of the discriminant matrix $D$ associated with the $P$ matrix defined in \eqref{eq:pmark}, which has the form
\begin{equation}
D=\begin{pmatrix}
1 & 0\\
0 & \frac{1}{N} \wt{e}\wt{e}^{T}
\end{pmatrix},
\label{eq:dmark}
\end{equation}
the success probability of measuring $\ket{0^n}$ can be much less than 1.

To see this, we recognize that the matrix $D$ defined in \eqref{eq:dmark} has two nonzero eigenvalues $1$ and $(N-1)/N=1-\delta$, with the corresponding eigenvectors $\ket{\wt{\pi}} = (1,0,...,0)^{T}$ and $\ket{\wt{v}}=\frac{1}{\sqrt{N-1}}(0,1,1\ldots,1)^{T}$, respectively. 
Because
\begin{equation}
\ket{\psi_0}=\frac{1}{\sqrt{N}}\ket{0^n}\ket{\wt{\pi}}+\sqrt{\frac{N-1}{N}}\ket{0^n}\ket{\wt{v}},
\end{equation}
it follows from the fact that $U_D^{k}$ block encodes $T_k(D)$ that
\begin{equation}\label{eqn:tmp_OZk_eval}
U_D^k\ket{\psi_0}=\frac{1}{\sqrt{N}}\ket{0^n}T_k(1)\ket{\wt{\pi}}+\sqrt{\frac{N-1}{N}}\ket{0^n}T_k(1-\delta)\ket{\wt{v}}+\ket{\perp},
\end{equation}
where $\ket{\perp}$ is an unnormalized state satisfying $(\ket{0^n}\bra{0^n})\otimes I_n \ket{\perp}=0$. 
Since $T_k(1)=1$ for all $k$, we have
\begin{equation}
p(\ket{0^n})=\frac{1}{N}+\left(1-\frac{1}{N}\right)T^2_k(1-\delta).
\end{equation}
Using the fact that $T_k(1-\delta)=\cos(k\arccos(1-\delta))$, we can show that $T_k(1-\delta)\approx 0$ when $k$ satisfies
\begin{equation}
k\approx \frac{\pi}{2\arccos(1-\delta)}\approx \frac{\pi}{2\sqrt{2\delta}}=\frac{\pi\sqrt{N}}{2\sqrt{2}}.
\label{eq:kopt}
\end{equation}
Therefore, after taking $k=\lceil\frac{\pi\sqrt{N}}{2\sqrt{2}} \rceil$ steps of the quantum walk, 
the probability of successfully measuring $\ket{0^n}$ drops down to $1/N$, which is significantly
lower than $1$. This will allow us to declare, with high confidence, the presence of a marked vertex 
in the graph in $\Or (\sqrt{N})$ steps of a quantum walk, which is a significant (quadratic) 
speed up compared to $\Or (N)$ steps of a random walk required in the classical setting for 
large $N$.

\begin{figure}[htp]
    \centering
    \subfloat[$U_D$ for \eqref{eq:pnomark}]{
        \includegraphics[height=3cm]{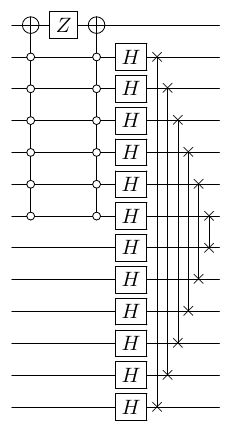}
    }
    \subfloat[$U_D$ for \eqref{eq:pmark}]{
        \includegraphics[height=3cm]{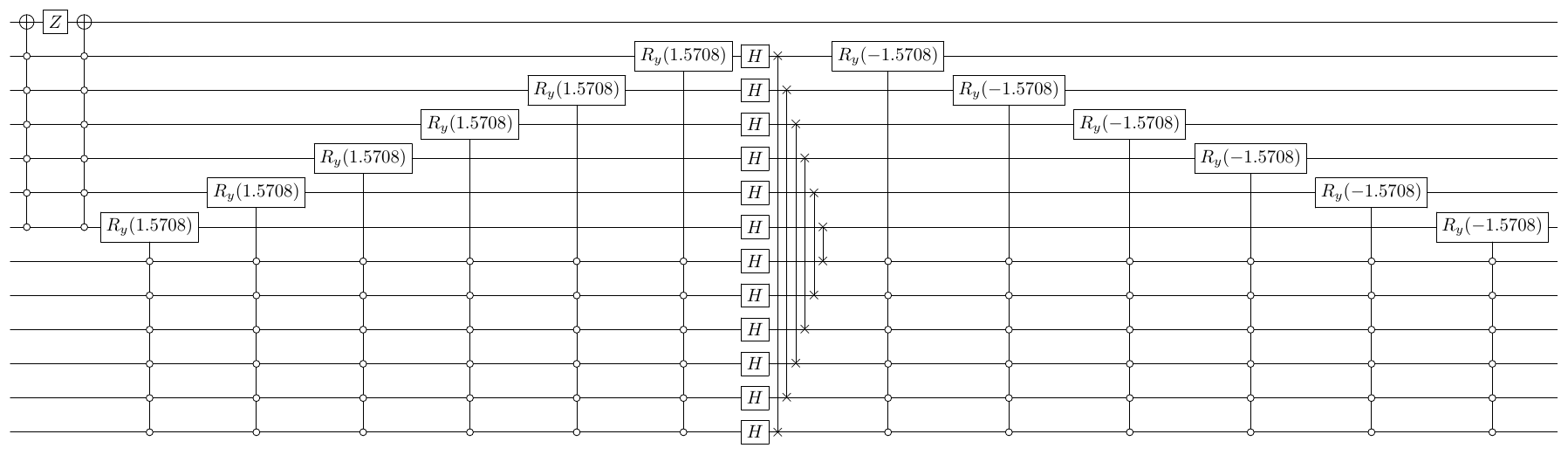}
    }
    \caption{Quantum circuits for block encoding of \eqref{eq:pnomark} and \eqref{eq:pmark}.}
    \label{fig:ble-circuits}
\end{figure}

We simulate this experiment for $n=6$ qubits in QCLAB. The script to reproduce this simulation is made
available on \url{https://github.com/QuantumComputingLab/qclab}. We first generate quantum circuits
that block encode the discriminant matrix for \eqref{eq:pnomark} and \eqref{eq:pmark}. These circuits
are given in \Cref{fig:ble-circuits}.

\begin{figure}[htp]
    \centering
%
%
\definecolor{mycolor1}{rgb}{0.00000,0.44700,0.74100}%
\definecolor{mycolor2}{rgb}{0.85000,0.32500,0.09800}%
\begin{tikzpicture}

\begin{axis}[%
scale only axis,
xmin=0,
xmax=30,
xlabel={$k$},
ymin=0,
ymax=1.1,
ylabel={Success probability$p(\ket{0^n})$},
title style={font=\bfseries},
title={Detecting a marked vertex ($n = 6$)},
legend style={legend cell align=left, align=left}
]
\addplot [color=mycolor1, mark=asterisk, mark options={solid, mycolor1}]
  table[row sep=crcr]{%
1	0.999999999999997\\
2	0.999999999999995\\
3	0.999999999999993\\
4	0.999999999999992\\
5	0.99999999999999\\
6	0.999999999999988\\
7	0.999999999999986\\
8	0.999999999999984\\
9	0.999999999999982\\
10	0.99999999999998\\
11	0.999999999999978\\
12	0.999999999999976\\
13	0.999999999999974\\
14	0.999999999999973\\
15	0.999999999999971\\
16	0.999999999999969\\
17	0.999999999999967\\
18	0.999999999999965\\
19	0.999999999999963\\
20	0.999999999999961\\
21	0.999999999999959\\
22	0.999999999999958\\
23	0.999999999999956\\
24	0.999999999999954\\
25	0.999999999999952\\
26	0.99999999999995\\
27	0.999999999999948\\
};
\addlegendentry{D}

\addplot [color=mycolor2, dashed, mark=o, mark options={solid, mycolor2}]
  table[row sep=crcr]{%
1	0.969478607177732\\
2	0.881699796766039\\
3	0.747550198571839\\
4	0.583667506902421\\
5	0.410377016525379\\
6	0.249170809743333\\
7	0.12004223446773\\
8	0.039006260483726\\
9	0.0161132478459778\\
10	0.0542024666765494\\
11	0.148549961124383\\
12	0.287454430608547\\
13	0.453688465339499\\
14	0.626635148900414\\
15	0.784845039091841\\
16	0.908696401485742\\
17	0.982828764378967\\
18	0.998047977295502\\
19	0.952466501133393\\
20	0.85173750725259\\
21	0.708353751730105\\
22	0.540098180807498\\
23	0.367838428769115\\
24	0.212938742205654\\
25	0.0946103127123874\\
26	0.0275286388690463\\
27	0.0200134204589353\\
};
\addlegendentry{Dt}

\addplot [color=black, dotted]
  table[row sep=crcr]{%
9	0\\
9	1.1\\
};
\addlegendentry{$\text{k}_{\text{opt}}$}

\end{axis}

\end{tikzpicture}%
    \caption{Simulation results for detecting the marked vertex.}
    \label{fig:simulation}
\end{figure}
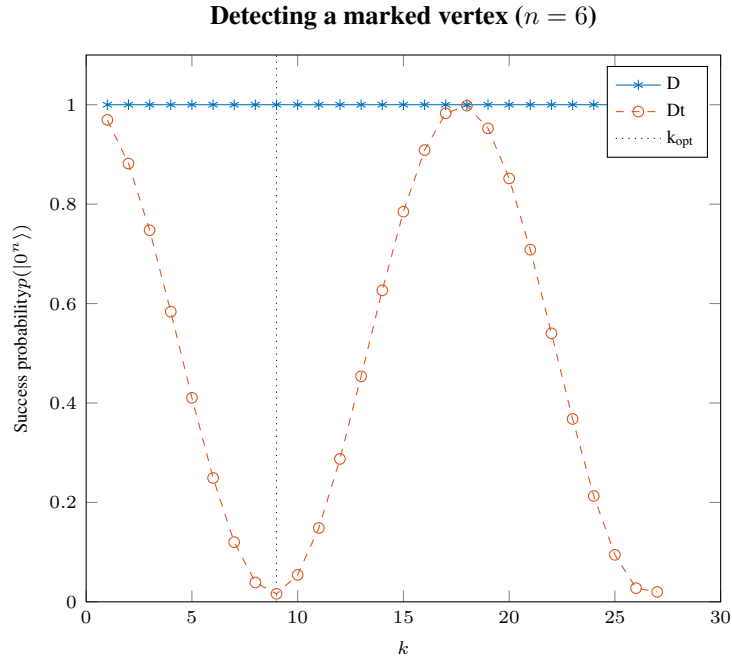

Starting from the state \eqref{eq:psi0}, we repeat both circuits for $k$ steps and measure 
the success probability \eqref{eq:prob}. The results of this simulation are shown in \Cref{fig:simulation}.
We observe that  it is sufficient to use a very small $k$ to distinguish the two states, and
 the two states are maximally distinguishable for $k_{opt}$ given by \eqref{eq:kopt} 
as indicated by the dotted line. 

%
%
%

\FloatBarrier

\bibliographystyle{unsrtnat}
\bibliography{references}

\end{document}